\documentclass[twocolumn,aps,prb,showpacs]{revtex4}
\usepackage{epsfig}
\usepackage{dcolumn}
\usepackage{graphicx}
\usepackage{amsmath}
\usepackage{amsfonts}
\usepackage{bm}
\usepackage{amssymb}
\usepackage{array}
\usepackage{color}

\newcommand{\vare}{\varepsilon}

\renewcommand{\Vec}[1] {{\bm #1}}

\def\k { {\Vec{k}} }
\def\p { {\Vec{p}} }
\def\q{ {\Vec {q}} }
\def\Q{ {\Vec {Q}} }

\def\mp { {\mathbf{p}} }

\begin{document}
\title{Spin resonance in AFe$_2$Se$_2$ with $s$-wave pairing symmetry}

\author{S. Pandey$^1$}
\author{A. V. Chubukov$^2$}
\author{M. Khodas$^1$}

\affiliation{$^1$Department of Physics and Astronomy, University of Iowa, Iowa City, Iowa 52242, USA \\
$^2$Department of Physics, University of Wisconsin, Madison, Wisconsin 53706,
 USA}


\begin{abstract}
We study spin resonance in the superconducting state of recently discovered alkali-intercalated iron selenide materials  A$_x$Fe$_{2-y}$Se$_2$ (A = K, Rb, Cs) in which the Fermi surface has only electron pockets. Recent angle- resolved photoemission spectroscopy (ARPES) studies [M. Xu et al., Phys. Rev. B 85, 220504(R) (2012)] were interpreted as strong evidence for s-wave gap in these materials, while the observation of the resonance peak in neutron scattering measurements [G. Friemel et al., Phys. Rev. B 85, 140511 (2012)] suggests that the gap must have different signs at Fermi surface points connected by the momentum at which the resonance has been observed. We consider recently proposed unconventional $s^{+-}$ superconducting state of  A$_x$Fe$_{2-y}$Se$_2$ with superconducting gap changing sign between the hybridized electron pockets. We argue that such a state supports a spin resonance. We compute the dynamical structure factor and show that it is consistent with the results of inelastic neutron scattering.
\end{abstract}

\maketitle

\section{Introduction}
\label{Sec:Intro}

Since its discovery \cite{Kamihara2008} the superconductivity in iron-based compounds remains one of the most active research frontiers for the past few years \cite{Mazin2010,Paglione2010,Johnston2010,Stewart2011,Basov2011}.
Of particular importance is the understanding of the microscopic mechanisms of superconductivity in these materials.
The iron-based SCs
are multi-band materials with conduction bands derived from iron $d$ orbitals and pnictide $p-$orbitals \cite{Mazin2008a,Mazin2009}.
The Fe sublattice has a simple tetragonal form
with 1 atom per unit cell, and the corresponding Fe-only Brillouin zone (BZ) is a  rectangular parallelepiped.
Throughout the paper we will refer to Fe-only BZ as 1FeBZ or, equivalently, unfolded BZ.
According to both Angle Resolved Photoemission (ARPES) \cite{Ding2008,Wray2008} and density functional theory (DFT), most of Fe-pnictides have a quasi two-dimensional band structure with two hole pockets centered at the $\Gamma$-point, and two electron pockets at $(0,\pi)$ and $(\pi,0)$ in 1FeBZ.
 In some systems, there is an additional 3D hole FS near $p_z =\pi$ and $(p_x,p_y) = (\pi,\pi)$.  

It is widely believed that in most Fe-pnictides superconducting order parameter (OP) has $s^{+-}$ symmetry \cite{Mazin2008,Kuroki2008,Hirschfeld2011,Chubukov2012}.
Such an OP changes sign between the hole and electron pockets and has a full lattice symmetry.
The inelastic neutron scattering experiments done on these systems revealed a spin resonance peak with the largest intensity at the neutron scattering momentum close to $(0,\pi)$ in 1FeBZ
 \cite{Christianson2008,Lumsden2009,Inosov2010}.
The spin resonance in FeSCs can be explained naturally within the $s^{+-}$ scenario, because
 $(0,\pi)$ and $(\pi,0)$ are momenta separating electron and hole pockets at which $s^{+-}$ gap has opposite signs~\cite{Korshunov2008,Maier2011,Maiti2011a}.

This paper focuses on superconductivity in recently discovered iron selenides A$_x$Fe$_{2-y}$Se$_2$ (AFe$_2$Se$_2$)  intercalated by an alkali metal, A = K,Rb,Cs \cite{Dagotto2013}.
These
 superconductors
 with $T_c \simeq 30$K \cite{Guo2010,Wang2011b,Ying2011} are isostructural with 122 family of Fe-pnictides.

Selenides differ from pnictides by a pronounced normal state transport anomalies and the presence of iron vacancies.
Superconductivity in AFe$_2$Se$_2$ is present simultaneously with local spin magnetism \cite{Liu2011}, but the two are very likely separated into spatially distinct domains.
 Several studies suggest that the superconductivity exists in stoichiometric domains without magnetic moments\cite{Chen2011a,Li2012,Li2012a}, while iron-vacancies are concentrated  in magnetic domains where they order~\cite{Wang2011c,Luo2011,Bosak2012}.
Although the  exact relationship between the magnetism and superconductivity is not yet settled,
 we believe there is enough evidence to separate superconductivity from local magnetism
   and consider
    superconductivity within  an effective itinerant low-energy model, without Fe vacancies.

\begin{figure}[t]
\begin{center}
\includegraphics[width=1.0\columnwidth]{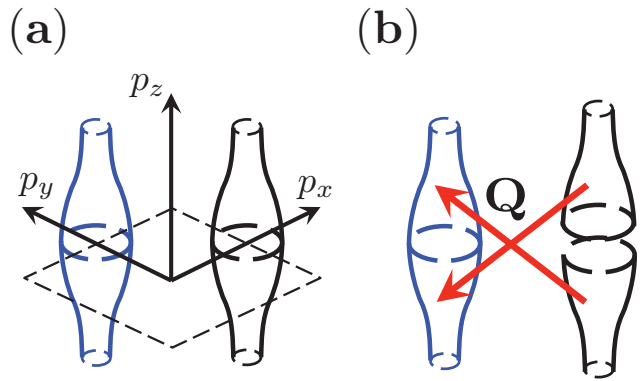}
\caption{(color online) (a) Schematic representation of the two electron pockets in unfolded 1FeBZ.
One pocket (blue) is
  centered along the $(0,\pi,p_z)$ vertical line and the other (black) is
  centered along the $(\pi,0,p_z)$  line.
Both Fermi surfaces are bounded from top and bottom by $p_z = \pm \pi$.
The 1FeBZ boundary crosses the $p_z=0$ along the thick solid (black) line.
(b) The three-dimensional folding specific to 122 systems with tetragonal body-centered crystal
 structure~\cite{Nekrasov2008,Su2009,Guo2010,Mazin2011}.
The thick solid (red) arrow denotes the folding vector $\Vec{Q} = (\pi,\pi,\pi)$.
This vector connects, in particular, the points $(\pi,0,0)$ and $(0,\pi,\pi)$.
The folding by $\Vec{Q}$ can be understood as if one pocket is cut in two along the $p_z=0$ plane, and the two halves are displaced by a vector $\Q$ in such a way that  the upper(lower) half is clipped underneath (above) the $p_z=0$ plane.}
\label{f1}
\end{center}
\end{figure}

Unlike in pnictides, where the Fermi surface has both electron and hole pockets, in selenides only electron pockets are present, according to ARPES \cite{Zhang2011,Qian2011,Wang2011,Zhao2011,Xu2012}.
The two largest Fermi pockets are centered at $(0,\pi)$
and $(\pi,0)$ in XY plane, and evolve as functions of $p_z$ (see Fig.~\ref{f1}(a).
 Hole pockets are lifted by about $60 meV$ from the FS\cite{Qian2011}. ARPES studies~\cite{Qian2011,Xu2012}
 found an additional 3D electron pocket centered at $p_z =\pi$ and at $p_x = p_y=0$.

Because hole pockets are absent, the  conventional scenario for $s^{+-}$ superconductivity  due to interaction between low-energy fermions near electron and hole pockets is questionable. It has been listed as possible explanation of the data\cite{Hirschfeld2011} (and termed as the
 ``incipient'' $s^{+-}$ order),  however because hole states are gapped,
 $T_c$ for such oder comes out  noticeably lower than in Fe-pnictides \cite{Hirschfeld2011}, in disagreement with the data.

Several alternative scenarios have been proposed, with the emphasize on the interaction between electron pockets, potentially enhanced by magnetic fluctuations at momentum separating the two electron pockets (i.e., at momentum $(\pi,\pi)$ in 1FeBZ). Strong inter-pocket interaction is necessary to overcome intra-pocket repulsion.
 Two scenarios propose a conventional pairing of fermions with momenta ${\p}$ and $-{\p}$  on one electron pocket due to interaction with fermions near the other pocket.  One proposal\cite{Maier2011a,Das2011,Das2011a,Wang2011c,Maiti2011b,Maiti2011c} is that inter-pocket interaction is strong and repulsive. In this case,
  the system develops a superconducting order in which the gap changes sign between the two electron pockets.
Such a gap necessarily has $d-$wave symmetry because it changes sign under the rotation from $X$ to $Y$ axis.
Another proposal~\cite{Yu2011b,Fang2011} is that inter-pocket interaction is strong and attractive.
This happens when, e.g.,  the underlying microscopic
 model is taken as the itinerant version of $J_1-J_2$ model with spin-spin interaction.
Then a superconducting gap does not change sign between electron pockets, i.e., superconducting state is a conventional $s-$wave.

Each of the two scenarios agrees with some experiments and disagrees with the others. A near-constant gap has been observed on a small 3D electron pocket centered at
 Z -point ($p_z = \pi$, $p_x = p_y =0$).
Taken at a face value (i.e., assuming that this is not a surface effect),
  this result is consistent with $s-$wave gap and rules out $d-$wave.  On the other hand, a spin resonance has been observed below $T_c$ in inelastic
  neutron scattering experiments~\cite{Park2011,Friemel2012,Friemel2012a,Taylor2012,Wang2012}.
If the resonance mode is a spin-exciton, as it is believed to be the case in Fe-pnictides and other unconventional superconductors~\cite{Eschrig2006},
  it requires a sign change of the gap.  The observation of the resonance then rules out a
  conventional sign-preserving $s-$wave and was interpreted as an argument for a d-wave gap~\cite{Maier2011a,Maier2012}.

\begin{figure}[t]
\begin{center}
\includegraphics[width=0.95\columnwidth]{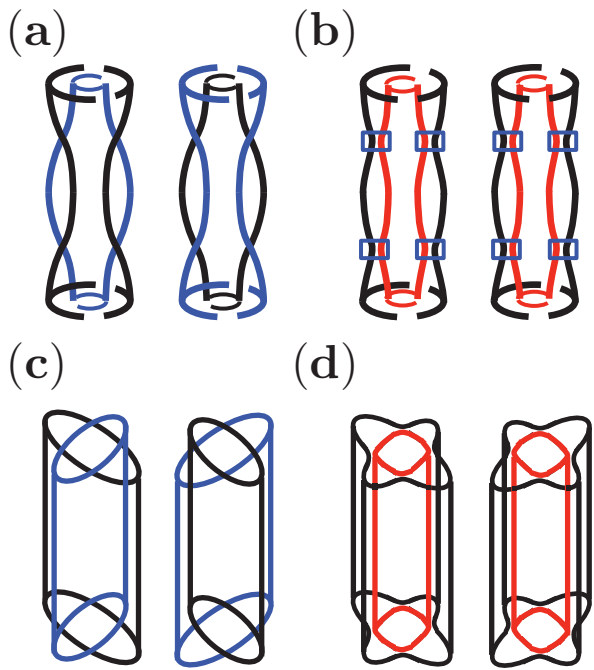}
\caption{(color online)
Fermi pockets in the the folded representation.
Panels (a) and (c) show the result of the folding without actual hybridization for the cases of strong and weak $p_z$ dispersion, respectively.
In each figure, one pair of Fermi surfaces is centered at $(\pi,\pi)$ in the XY plane, the other at $(\pi,-\pi)$.
The folding without  hybridization results in the two Fermi pockets in the corners of folded BZ, which overlap either only at a particular $|p_z|=\pi/2$, in the case of strong dispersion (panel a), or along vertical lines in the case of weak dispersion (panel c).
 For strong dispersion, the two Fermi surfaces in each cross-section at a given $p_z$ are elliptical, except for $|p_z|=\pi/2$,
  where they are near-circular
(more precisely, $C_4$ symmetric).
The long axis of a cylinder rotates by $90^{\circ}$ between $p_z =0$ and $p_z = \pi$,
 see Fig.~\ref{fig_insert}.
 For weak dispersion, the crossed ellipses in each cross-section are the same for all $p_z$.
Panels (b) and (d) show the Fermi surfaces in the presence of a finite hybridization, again for strong and weak $p_z$ dispersion.
A finite hybridization lifts the degeneracy, and the crossing lines are eliminated.
For strong dispersion (panel b)  the hybridization affects mostly the regions framed by (blue) rectangles.
 The actual hybridized Fermi surfaces are shown in Fig.~\ref{fig_insert}.
   For weak $p_z$ dispersion (panel d), the hybridization affects the region where the two pockets in
panel (c) cross.
The hybridized Fermi surfaces are again two cylinders, one inside the other, but now each is $C_4$ symmetric in every cross-section.
The smaller one is nearly a circular cylinder, the larger one has a substantial anisotropy in the XY plane.}
\label{f1_fold}
\end{center}
\end{figure}

\begin{figure}
\begin{center}
\includegraphics[width=1.0\columnwidth]{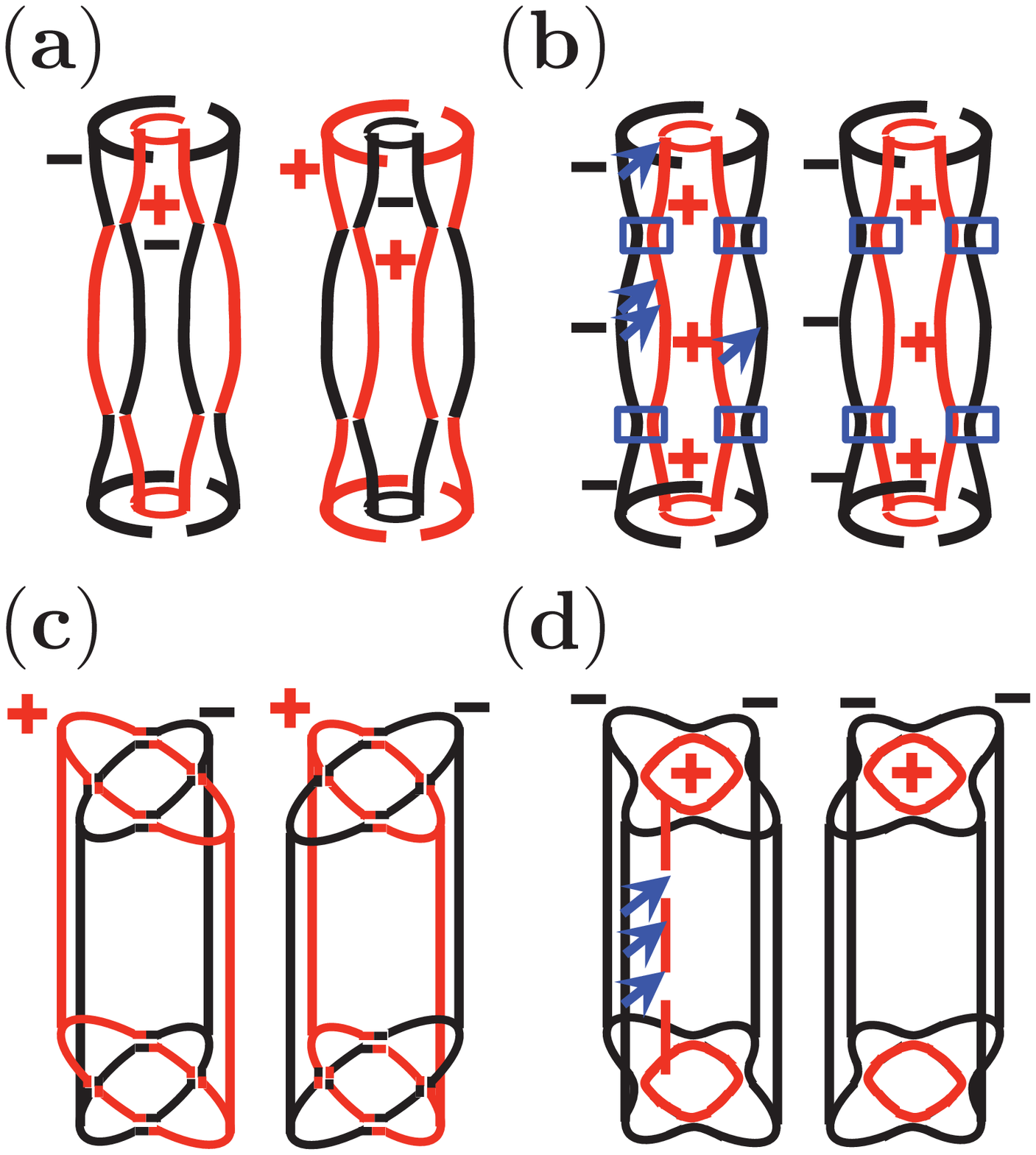}
\caption{(color online)
Superconducting gap on the folded Fermi surfaces.
 Panels (a) and (c) -- a $d-$wave state.
 For strong $p_z$ dispersion (panel a),  the gap has opposite sign on the two Fermi surfaces in each cross-section and
 changes sign along each Fermi surface upon varying $p_z$.
 As a result the magnitude of the gap vanishes for particular $p_z$ (horizontal nodes).
 For weak $p_z$ dispersion (panel c), the gap has $\cos 2 \phi$ structure with nodes on each of the two Fermi surfaces in every cross-section.
 In this limit, nodal lines are vertical.
 Panels (b) and (d) --- an $s^{+-}$ state.
For both weak and strong $p_z$ dispersion, the gap  changes sign between the bonding (inner, red) and anti-bonding (outer, blue) Fermi
 surfaces, but preserves its sign along each Fermi surface in every cross-section and does not change sign as a function of $p_z$.
}\label{f2}
\end{center}
\end{figure}

 There exists, however, another problem with the $d-$wave state,  even if we forget momentarily about ARPES measurements on the $Z$-pocket.
 Namely, specific heat and other data on AFe$_2$Se$_2$ show~\cite{Zhang2011,Mou2011} that there are no nodes in the superconducting gap.  In a given 2D cross-section, $d-$wave
   state due to repulsion between electron pockets yields a ``plus-minus'' gap, which is seemingly nodeless. However, the size and orientation of the two electron pockets in 122-type structures vary with $p_z$,
   (see Fig.~\ref{f1_fold})
   and one can verify (see Fig.~\ref{f2}(a)) that the ``plus'' and ``minus'' gaps necessarily cross at some $p_z$. Around this $p_z$, the hybridization between the two pockets, caused by the presence of a pnictide either above or below Fe plane, splits the two pockets into bonding and anti-bonding states.  One can show quite generally (see Refs.~[\onlinecite{Mazin,Mazin2011,Khodas2012a}] and Fig.~\ref{f2}(a,c)) that the gap on each hybridized Fermi surface evolves from ``plus'' to ``minus'' and must necessarily have
nodes, in disagreement with the data.
\begin{figure}[h]
\begin{center}
\includegraphics[width=0.85\columnwidth]{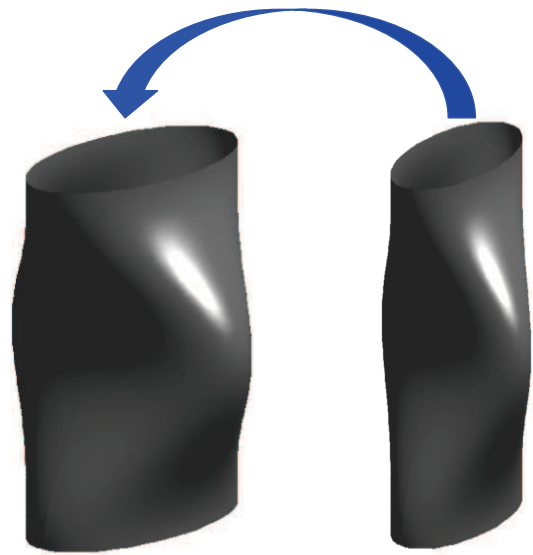}
\caption{
The 3D hybridization in 122 systems with tetragonal body-centered crystal structure, in the limit of strong $p_z$ dispersion.
The two warped Fermi surfaces are shown separately, but the smaller one is actually inside the larger one, as the arrow indicates.
Each Fermi surface is a corrugated elliptical cylinder with a near-circular cross-section at $p_z \approx \pm \pi/2$ (more precisely, $C_4$ symmetric cross-section).
The long axis of each cylinder is rotated by $90^{\circ}$ between $p_z =0$ and $p_z =\pi$.
}
\label{fig_insert}
\end{center}
\end{figure}

There exists a third scenario~\cite{Mazin2011,Khodas2012a},
 which alleviates the contradiction between ARPES and neutron scattering data and is consistent with the measurements which show a no-nodal gap.
  Namely, the same interaction which gives rise to a ``plus-minus'' d-wave state in which Cooper pairs are made out of fermions
  on the same pocket also gives rise to an s-wave state in which pairing at least partly involves pairing between fermions belonging to different pockets.
   This ``other'' s-wave state is best understood once one converts to the actual (physical) BZ with two Fe atoms in the unit cell (2FeBZ) and includes the hybridization
 between the pockets, which splits them into bonding and anti-bonding Fermi pockets which we will label as $a$ and $b$.  The ``other'' s-wave gap remains roughly constant along each pocket after hybridization, but changes sign between them,  $\mathrm{sgn}(\Delta_a) = -\mathrm{sgn}(\Delta_b)$.

We recall that the hybridization in 122 compounds can be traced to the checkerboard arrangement of
pnictogen/chalcogene atoms staggered above and below the iron planes, \cite{Calderon2009,Carrington2009,Coldea2010,Khodas2012}.
The iron lattice sites at
$i a \hat{x} + j a \hat{y} + k c \hat{z}$,
with integer $i,j,k$ then belong to even and odd sublattices, defined by an even and odd $i+j+k$, respectively.
Because sublattices are inequivalent, the correct BZ is
the folded 2FeBZ, and in the folded zone the momenta $\Vec{p}$ and $\Vec{p}+\Vec{Q}$, where $\Vec{Q}$ is folding vector,  are equivalent.
The folding vector is  $\Vec{Q}=(\pi,\pi,0)$ in simple tetragonal systems such as 11 and 1111 materials, and $\Vec{Q} = ( \pi,\pi, \pi)$ in 122 materials with body-centered tetragonal crystal structure, like in AFe$_2$Se$_2$.

\begin{figure}
\begin{center}
\includegraphics[width=1.0\columnwidth]{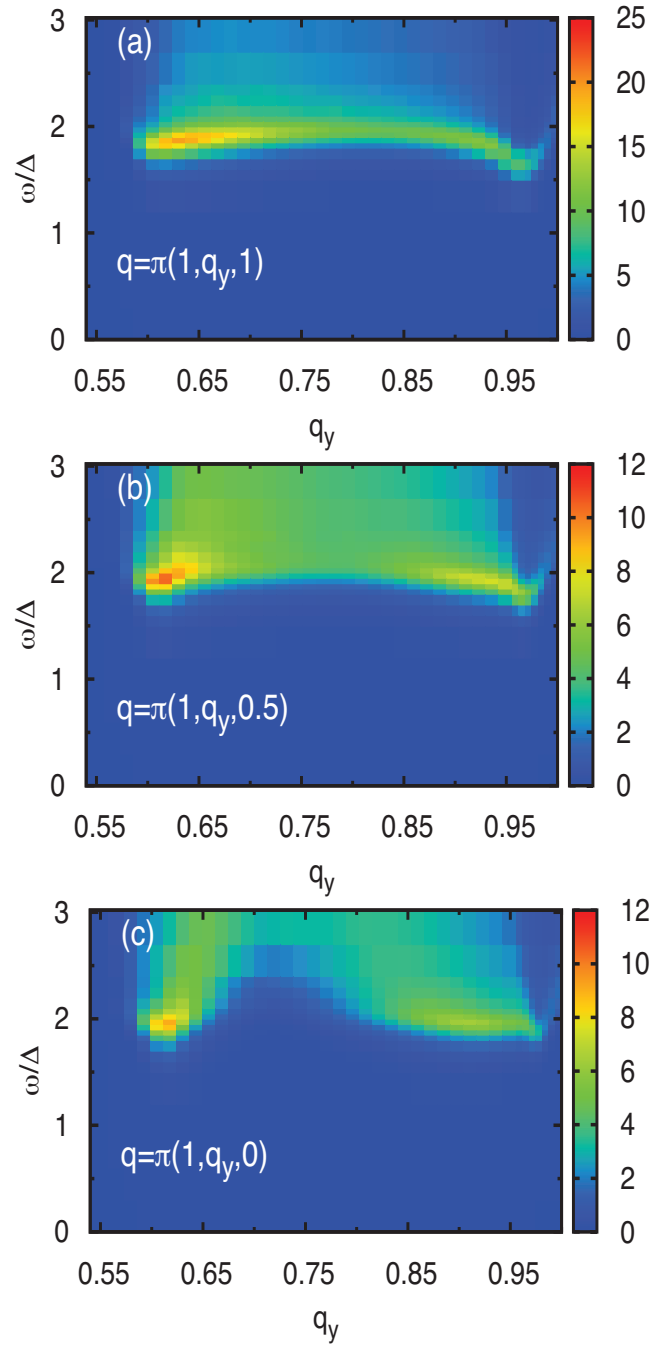}
\caption{(color online)
The color-plot of
 the dynamic structure factor
$S(\q,\omega)$ of an $s^{+-}$ superconductor for {\it weak} out-of-plane $p_z$-dispersion, see Fig.~\ref{f2}(d).
 To represent the weak dispersion limit, we set $\Lambda=0.1$ in Eqs.~\eqref{eqH},\eqref{dispersion}.
The $S(\q,\omega)$ is shown as a function of $q_y$
 (horizontal axis) and frequency $\omega$ (vertical axis) for a fixed $q_x = \pi$ at three different values of $q_z$
(a) $q_z= \pi$; (b) $q_z = \pi/2$; (c) $q_z = 0$.
The hybridization is
set to $\lambda=5$meV and the gap is $\Delta =10$meV.
The in-plane ellipticity is $\epsilon=0.1$.
A small imaginary part, $\Gamma = 1$meV was added to the frequency $\omega$ for regularization
of the numerical computation.}
\label{f4-color}
\end{center}
\end{figure}

\begin{figure}
\begin{center}
\includegraphics[width=1.0\columnwidth]{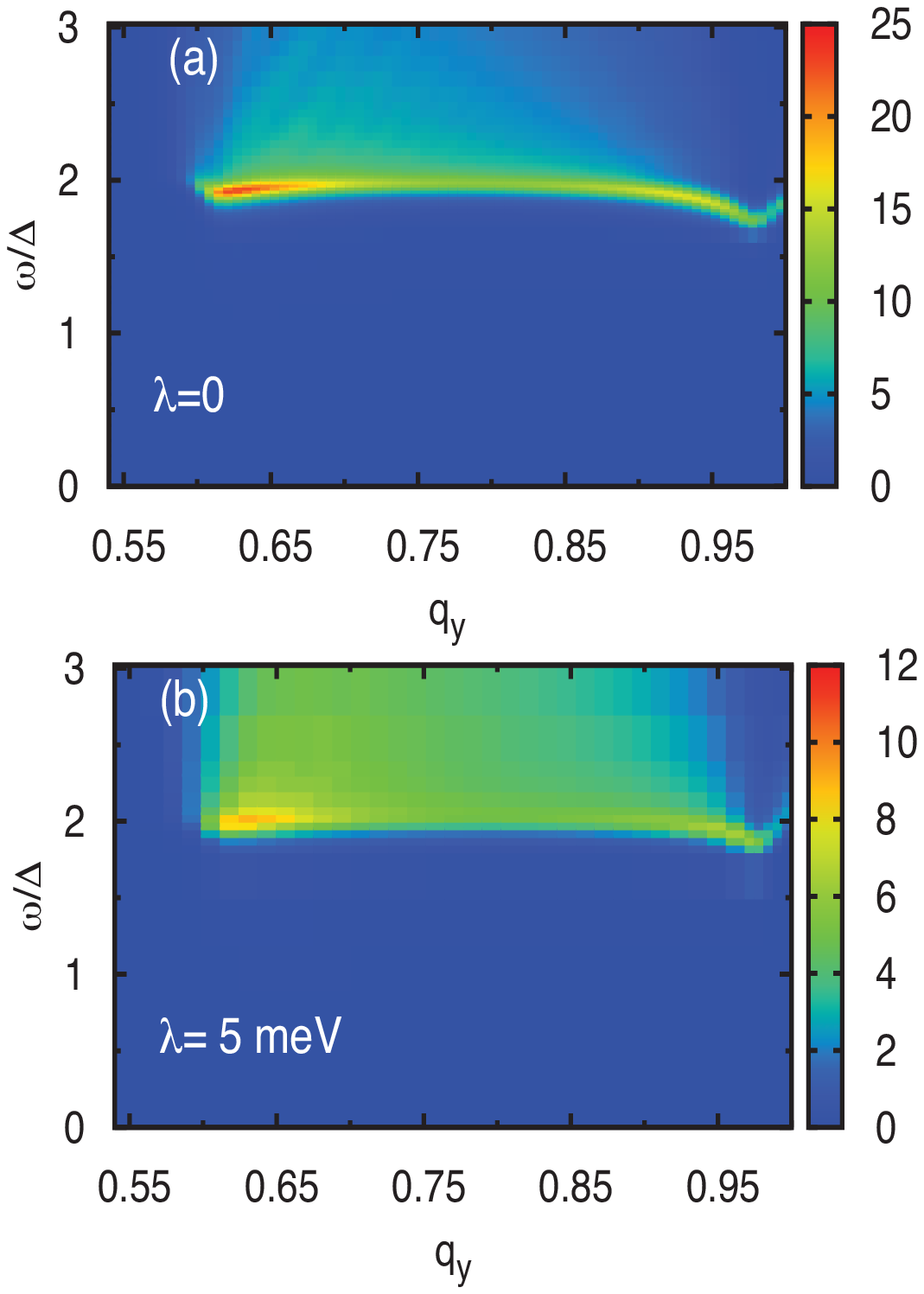}
\caption{(color online)
The color-plot of
the dynamic structure factor $S(\q,\omega)$ of an $s^{+-}$
superconductor for {\it strong} out-of-plane $p_z$-dispersion, see Fig.~\ref{f2}(b).
In contrast to Fig.~\ref{f4-color}, the ellipticity now changes sign at $|p_z| = \pi/2$.
The $S(\q,\omega)$ is shown as a function of $q_y$ (horizontal axis) and frequency $\omega$ (vertical axis) for fixed $q_x = \pi$, $q_z = \pi/2$ for (a) no hybridization, ($\lambda = 0$) and (b) $\lambda = 5$meV.
The superconducting gap is $\Delta = 10$meV.
A small imaginary part, $\Gamma = 0.5$meV was added to the frequency $\omega$  for regularization.
}
\label{f5-color}
\end{center}
\end{figure}

This ``other''  $s^{+-}$ state  is nodeless and in this respect is consistent with ARPES and other measurements which show that the gap likely has no nodes.
 A seemingly similar state can be obtained if one still assumes that the pairing is solely between $\p$ and $-\p$
 from the same pocket in the unfolded BZ, but
   the gap is higher-angular momentum $s-$wave state with  $\Delta (\phi) = \pm \Delta \cos{2\phi}$, where $\Delta$ is the angle along the FS counted from, say, $x-$axis, and plus and minus are for one or the other electron pocket. After folding and hybridization, this state also becomes $s^{+-}$, with the sign change of the gap
    between bonding and anti-bonding Fermi surfaces.  However, the gap still vanishes along the directions $\phi = \pm \pi/4$, at which $\cos 2 \phi =0$. This, again, is in contradiction with the data.

The goal of this paper is to demonstrate that the ``other'' $s^{+-}$ state, proposed for AFe$_2$Se$_2$  is not only nodeless $s-$wave state, but is also consistent with the observation of a spin resonance in the inelastic neutron scattering.

The paper is organized as follows.
In the next Section we present qualitative reasoning and summarize our results for a reader not interested in technical details.
 Sec.~\ref{sec:Low} we
 introduce the
  low energy model, set up the formalism for
   the
   analysis of
   the spin susceptibility, and
  discuss the ``other'' $s^{+-}$ superconducting state.
  In Sec. ~\ref{sec_results} we present the results for the spin structure factor of this $s^{+-}$ superconductor. We first discuss, as a warm-up, the
artificial limit of zero hybridization and then discuss the actual case when the hybridization is finite (and strong enough to
 favor the $s^{+-}$ state over the $d-$wave state).
 We present our conclusions in Sec.~\ref{sec:conclusions}.

\section{Qualitative consideration and a brief summary of the results}

\subsection{Qualitative consideration}

Naively, the spin resonance is inevitable in the presence of the sign-changing OP.
The reasoning is that for sign-changing OP, superconductivity simultaneously gives rise to two features in the spin response: (i)
it gives rise to a gap $2\Delta$ in the spin excitations spectrum and (ii) spin component of the residual interaction between fermions is attractive.
The combination of these  two conditions gives rise to the excitonic resonance below $2\Delta$.  The residue of the resonance peak at momentum between bonding and anti-bonding Fermi surfaces
 is proportional to
the spin coherence factor, $(1 - \Delta_a \Delta_b / |\Delta_a| |\Delta_b|)$, and the latter is non-zero if  the OP has opposite sign on bonding ($a$) and anti-bonding ($b$) bands.
However, this condition is {\it necessary} but not {\it sufficient}.
To see this, neglect momentarily the ellipticity of electron pockets and the $p_z$ dispersion, i.e., approximate each pocket by a circular cylinder.
 Bonding and anti-bonding states are then the sum and the difference of the states of the original (non-hybridized fermions). In operator notations,
 $a_{\Vec{p}} = (\beta_{1,\Vec{p}} + \beta_{2,\Vec{p}+\Vec{Q}} )/ \sqrt{2}$, and
$b_{\Vec{p}} = (\beta_{1,\Vec{p}} - \beta_{2,\Vec{p}+\Vec{Q}} )/ \sqrt{2}$, where
$\Vec{p} \approx (0,\pi,p_z)$ and $\Vec{p}+\Vec{Q} \approx (\pi,0,p_z+\pi)$, and subindices $1$ and $2$ label electron pockets.
One can easily verify\cite{Mazin} that in real space bonding and anti-bonding states reside on even and odd Fe-sublattices  respectively, and do not overlap.
For that reason, the
spin operator has zero matrix elements between them, hence the residue of the resonance vanishes.
Another way to understand this argument is to note that the spin operator does not discriminate between the two original pockets before the hybridization,
 i.e., it is symmetric under the exchange $\beta_{1} \leftrightarrow \beta_{2}$.
Since bonding and anti-bonding states have opposite parity under this operation, the symmetric spin operator cannot induce transitions between them.

 The above argument, however, applies only to Fermi pockets in the form of circular cylinders.
In reality, the original pockets are not circular for a generic $p_z$, and
 moreover hybridization and folding in 122 materials is a complex process in a three-dimensional BZ,
 Fig.~\ref{f1_fold},\ref{fig_insert}.
We show that the proper folding procedure by a vector,
$\Vec{Q} = (\pi,\pi,\pi)$ combined with the full three-dimensional band dispersion leads to $s^{+-}$ state on bonding and anti-bonding Fermi surfaces, for which
 the residue of the spin resonance is non-zero.  One particular reason for the existence of the resonance is that the structure of the two Fermi surfaces in 2FeBZ
  is such that they strongly overlap only in a subset of points along $p_z$ axis. Inside this range (framed by rectangles in Fig.~\ref{f2}(b))
   hybridization separates bonding and anti-bonding states into
   even and odd
sublattice states with near-zero overlap and hence near-zero contribution to the resonance.
     However, in other regions of $p_z$, the two pockets appear split already before hybridization. For these $p_z$, the effect of hybridization is minimal (if, as we assume,  hybridization is not too strong to exceed the energy difference between two split bands), and in real space each state resides on even and odd sublattices. The overlapping between the two states is then strong and the condition that the gap
    changes sign between the two Fermi surfaces becomes not only necessary but also  sufficient for the resonance.
The same reasoning also holds for the case of cylindrical FSs in 1FeBZ (no $p_z$ dependence), but with ellipses rather than circles in the cross-section. Then again, the two electron pockets overlap only near particular $(p_x,p_y)$, and in this $\p$ range hybridization generates bonding and anti-bonding states residing on different sublattices.
However,
away from the overlapping region
 the original states from two electron pockets are already well separated, and hybridization
 does not constrain the  states to either even or odd sublattices.  In this situation, again, the sign change of the gap between the two Fermi surfaces becomes not only necessary but also  sufficient condition for the resonance.

\subsection{A brief summary of the results}

In the next two Sections we present a detailed account of our
calculation of the dynamical structure factor $S(q, \omega) \propto \chi^{''} (q, \omega)$. Here we give a brief summary of our result for a reader not interested in technical details.

\subsubsection{Weak dispersion}

We verified that in the limit of weak dispersion, the
  ellipticity of electron pockets in 1FeBZ is necessary for the existence of resonance, as
cylindrical pockets are strongly hybridized into bonding and anti-bonding states, which are not connected by the spin operator.
As a result the residue of the resonance peak vanishes.
In contrast, for finite ellipticity, a finite portion of the Fermi surface remains unaffected by hybridization.
The transitions between such states contribute to the spin resonance, as indicated by the arrows in Fig.~\ref{f2}(d).
Our numerical results in the weak dispersion limit are presented in Fig.~\ref{f4-color}.
We have found that the resonance mode becomes stronger with increasing pocket ellipticity.
The intensity of the resonance is maximized for neutron momenta $\Vec{q}$ such that the two Fermi pockets touch each other when one of them is shifted by a vector $\Vec{q}$ in a BZ.
The two distinct minima in Fig.~\ref{f4-color} refer to the external and internal touching conditions.
The large intensity at the minima  is due to the increased phase space for the two particle excitation at these particular wave-vectors, \cite{Maiti2011a,Maier2011a,Maier2012}.
The out-of-plane dispersion of the resonance mode is weak
 because pockets are weakly dispersive in the out-of-plane momentum $p_z$.

\subsubsection{Strong dispersion}

The representative plots of spin structure factor
  for the case of strong dispersion (see Fig.~\ref{f2}(b))
  are presented in Fig.~\ref{f5-color}.
In this case the phase space for the transitions which contribute to the spin structural factor
  is suppressed for $q_z = 0$ and is maximized for $q_z \approx \pi$.
 The minima at the two touching momenta in Fig.~\ref{f5-color} are less pronounced than in Fig.~\ref{f4-color}.
It is natural since the touching condition can be satisfied only approximately in the presence of strong $p_z$-dispersion
of the two Fermi surfaces.
 For the FS's as observed by ARPES in AFe$_2$Se$_2$ materials, the in-plane component of the external touching momentum is close to $(\pi,\pi/2)$.
  This is consistent with the momenta at which the  maximum intensity of neutron scattering has been observed
  in Rb$_x$Fe$_{2-y}$Se$_2$ \cite{Friemel2012}.

\section{Spin susceptibility in the presence of intra- and inter-pocket pairing}
\label{sec:Spin}
\subsection{Low energy model with inter-band hybridization. 1FeBZ formulation.}
\label{sec:Low}
We model the electronic structure of AFe$_2$Se$_2$ by a two band model with two electron-like Fermi pockets around $(0,\pi,p_z)$ and $(\pi,0,p_z)$ in the 1FeBZ.
The quadratic part of the Hamiltonian is
\begin{equation}
\label{eqH}
H_2 =
 \sum_{\p,\sigma} \left[
\vare^{\beta_{1}}_{\p} \beta_{1\p,\sigma}^\dag
\beta_{1\p,\sigma} +
\vare^{\beta_{2}}_{\p+\Q} \beta_{2\p+\Q,\sigma}^\dag \beta_{2\p+\Q,\sigma} \right],
\end{equation}
where $\beta_1$ and $\beta_2$ refer to the two electron bands and $\Q = (\pi,\pi,\pi)$.
We model in-plane and out-of-plane dispersions by
\begin{align}\label{dispersion}
\vare^{\beta_1}_{\p} = & -t (p_z) [\{1+\epsilon (p_z)\} \{ \cos (p_x) -1 \}
 \notag \\
& +
 \{1-\epsilon (p_z) \} \{ \cos (p_y + \pi ) -1 \} ] -\mu \, ,
  \notag \\
\vare^{\beta_2}_{\p}  = & -t (p_z) [\{1 - \epsilon (p_z)\} \{ \cos (p_x +\pi) -1 \}
\notag \\
& +
\{1+\epsilon (p_z)\} \{ \cos (p_y)  -1 \} ] -\mu \, ,
\end{align}
where $\epsilon(p_z)$ is the in-plane pocket ellipticity and
$t(p_z) = t \left[ 1- \Lambda \cos (p_z) \right]$.
The parameters $\Lambda$ and $\epsilon (p_z)$ control the $p_z$ dependence of the size and shape of the Fermi surfaces, respectively.
We choose them to reproduce the ellipticity and $p_z$ dispersion obtained for systems with AFe$_2$Se$_2$ composition (122-type structure)
in DFT calculations~\cite{Mazin2011}.

We describe the hybridization between the two pockets by
\begin{equation}\label{Hhybrid}
H_{hyb} = \lambda (\beta_{1\p,\sigma}^\dag \beta_{2\p+\Q,\sigma} + h.c )
\end{equation}
The hybridization term emerges because there are two non-equivalent positions of a chalcogen (Se for  AFe$_2$Se$_2$) above and below Fe plane, and the
 correct unit cell contains two Fe atoms (2FeBZ). Because of the doubling, there exist, in 1FeBZ, processes with momentum transfer $\Q = (\pi,\pi,\pi)$, i.e., the scattering processes in which a fermion near one electron pocket is annihilated, and a fermion near the other pocket is created.
 The hybridization parameter $\lambda$ has to be evaluated using a microscopic model for electron hopping and generally depends on the
  magnitude of the Fermi momentum (it vanishes for point-like Fermi surfaces) and on the angle
$\phi$
  along the pockets~\cite{Carrington2009,Coldea2010,Mazin2011,Khodas2012}.
   In the absence of spin-orbit coupling, $\lambda (\phi)$ vanishes along the diagonal directions $\phi = \pm \pi/4$, but $\lambda (\pi/4)$
    remains finite when spin-orbit interaction is included.  Our consideration and results do not depend qualitatively on the form of $\lambda (\phi)$
    and on whether or not it vanishes at $\pm \pi/4$. To simplify the discussion, we just set $\lambda (\phi)$ to be a constant $\lambda$.

\begin{figure}
\begin{center}
\includegraphics[width=1.0\columnwidth]{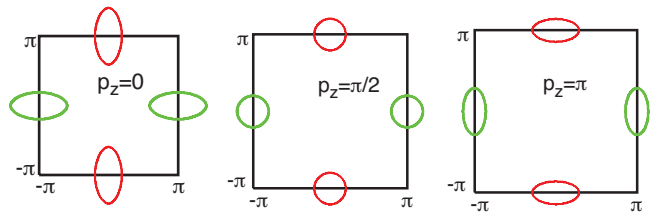}
\end{center}
\caption{$p_z$-variation of unhybridized Fermi surfaces,
as shown by taking the cuts at three different $p_z$.
The ellipticity changes sign at $|p_z|=\pi/2$, at which the pockets are $C_4$-symmetric.
The size of Fermi surfaces decreases with increasing $p_z$.
The band parameters used are $t=0.7 eV, \mu=0.14 eV,\Lambda=0.3, \epsilon=0.6$.}
\label{fig:FS}
\end{figure}

Below we separately analyze the two limiting cases of the weak and strong $p_z$ dispersion
(the cases presented in Figs.~\ref{f2}(d) and \ref{f2}(b), respectively).
The two limits are modeled in Eq.~\eqref{dispersion} by a constant  and a sign-changing ellipticity, $\epsilon (p_z)= \epsilon$ and $\epsilon (p_z)= \epsilon \cos (p_z)$, respectively.
Explicitly, $\epsilon (p_z)= \epsilon$ $\left[\epsilon (p_z)= \epsilon \cos (p_z)\right]$ describes the weak [strong] out-of-plane dispersion.
The constant $p_z$ cross sections of the Fermi pockets for the case of the strong dispersion are shown on Fig.~\ref{fig:FS}.
In  numerical calculations we used
$t=0.7 eV, \mu=0.14 eV, \Lambda=0.1, \epsilon=0.1$ (unless specified otherwise).

The interaction Hamiltonian involves both the intra-pocket and inter-pocket momentum-conserving
four-fermion interactions given by
\begin{align}
H_{int} &=  \frac {u_{1}}{2} \! \sum\! \left[{\beta}^{\dagger}_{1\p_3 \sigma}
 {\beta}^{\dagger}_{2\p_4 \sigma'}  {\beta}_{2\p_2 \sigma'}
 {\beta}_{1\p_1 \sigma} \! +\!
 \left( \beta_{1\p_i} \! \leftrightarrow \! \beta_{2\p_i} \right) \right]
  \notag \\
 + &
\frac{u_{2}}{2}\! \sum \left[{\beta}^{\dagger}_{2\p_3 \sigma}
 {\beta}^{\dagger}_{1\p_4 \sigma'}  {\beta}_{2\p_2 \sigma'}
{\beta}_{1\p_1 \sigma}\! +\! \left( \beta_{1\p_i}  \!\leftrightarrow  \!\beta_{2\p_i} \right)  \right] \notag \\
+ &
\frac{u_{3}}{2}\!  \sum \left[ {\beta}^{\dagger}_{2\p_3 \sigma}
 {\beta}^{\dagger}_{2\p_4 \sigma'}  {\beta}_{1\p_2 \sigma'}
  {\beta}_{1\p_1 \sigma} \! +\! \left( \beta_{1\p_i}  \!\leftrightarrow \! \beta_{2\p_i} \right) \right] \notag \\
 + &
\frac{u_4}{2}\! \sum \left[{\beta}^{\dagger}_{1\p_3 \sigma}
{\beta}^{\dagger}_{1\p_4 \sigma'}
{\beta}_{1\p_2 \sigma'} {\beta}_{1\p_1 \sigma} \!+\! \left( \beta_{1\p_i}  \!\leftrightarrow \! \beta_{2\p_i} \right) \right]\, .
 \label{eq:2}
\end{align}

There also exist interaction terms with momentum transfer ${\bf Q}$, but we earlier found~\cite{Khodas2012a} that they are not relevant for the pairing
 and can be omitted.

In the superconducting state, we truncate $H_2 + H_{hyb} + H_{int}$ to the effective mean field Hamiltonian $\hat {\cal H}_{MF}$ in Nambu space constructed of
 ${\hat \psi_{\p}} \equiv
[\beta_{1\p\uparrow},\beta_{1-\p \downarrow}^{\dag},\beta_{2\p+\Q \uparrow},\beta_{2 -\p-\Q \downarrow}^{\dag}]^{T}$.
The Hamiltonian $\hat {\cal H}_{MF}$ is
given by
\begin{align}\label{H-BCS}
{\hat {\cal H}}_{MF} (\p)  = \begin{bmatrix}
\vare^{\beta_1}_\p  &  \Delta^{\beta_1\beta_1}_\p & \lambda & \Delta^{\beta_1 \bar{\beta_2}}_\p   \\
\Delta^{\beta_1\beta_1}_\p & -\vare^{\beta_1}_\p & \Delta^{\beta_1 \bar{\beta_2}}_\p & -\lambda \\
 \lambda & \Delta^{\bar \beta_2 \beta_1}_\p & \vare^{\bar{\beta_2}}_\p  &  \Delta^{\bar \beta_2 \bar \beta_2}_\p  \\
\Delta^{\bar \beta_2 \beta_1}_\p & -\lambda & \Delta^{\bar \beta_2 \bar \beta_2}_\p & -\vare^{\bar \beta_2}_\p
\end{bmatrix} \, ,
\end{align}
Here the band index with a bar denotes the shift in momentum by the hybridization vector $\Q$,
$\bar{\beta}_{i\p} \equiv \beta_{i\p+\Q}$.
In the mean field Hamiltonian, Eq.~\eqref{H-BCS}, the intra-band gap functions, such as
$\Delta^{\beta_{1}\beta_{1}}_\p$, describe conventional zero-momentum pairing, while
the gap functions such as $\Delta^{\beta_1 \bar{\beta_2}}_\p$, describe inter-band pairing at the total  momentum $\Q$ of a pair.

The Matsubara Green's function is a $4$ by $4$ matrix,

\begin{align}\label{Green-BCS-def}
{\hat {\cal G}} (\p,i\omega_n) &= -\langle {\hat \psi_{\p}} {\hat \psi_{\p}}^{\dagger}\rangle_{\omega_n}
 \nonumber \\
= & \begin{bmatrix}
 G_{11} (\mp) &  F_{11} (\mp) &  G_{1{\bar 2}} (\mp) &  F_{1 {\bar 2}} (\mp)\\
 F_{11}(\mp)  & -G_{11}(-\mp) &  F_{1{\bar 2}} (\mp) & -G_{1{\bar 2}} (-\mp) \\
 G_{{\bar 2}1}(\mp)  & F_{{\bar 2} 1}(\mp) & G_{\bar 2 \bar 2} (\mp) &  F_{\bar 2 \bar 2} (\mp)\\
 F_{{\bar 2}1} (\mp) & -G_{{\bar 2}1} (-\mp) &  F_{\bar 2 \bar 2} (\mp) & -G_{\bar 2 \bar 2} (-\mp)
\end{bmatrix}
\nonumber \\
= & (i\omega_n {\hat I} - {\cal {\hat H}}_{MF})^{-1}\, ,
\end{align}
where we have used the notations $\langle A B \rangle_{\omega_n} = \int_0^{\beta} d \tau e^{i\omega_n \tau}\langle T_{\tau} A(\tau) B(0) \rangle $,
$\mp=(\p,i\omega_n)$, and $(\beta_1,\beta_2)=(1,2)$. For example, $G_{11} (\mp) = -\langle \beta_{1\p} \beta_{1\p}^{\dag}\rangle_{\omega_n}$ and
$G_{1 {\bar 2}} (\mp) = -\langle \beta_{1\p} {\bar \beta_{2\p} }^{\dag}\rangle_{\omega_n}$ etc.
 The functions  $G$ and $F$ represent the normal and anomalous Green's functions.
 Note that the inter-band propagators such as $G_{1 {\bar 2}} (\mp)$, $F_{1 {\bar 2}} (\mp)$ etc., which connect the two different bands with a momentum transfer $Q$,
vanish identically in the absence of the hybridization.

To study the spin resonance we consider generalized susceptibility
\begin{align}\label{generalized}
\chi_{ijkl} (\q',\q'')= &
\int_0^{\beta} d \tau e^{i\Omega_m \tau}\langle T_{\tau} S_{ji}^{+}(\q',\tau) S_{kl}^{-}(-\q'',0) \rangle
\, ,
\notag \\
S_{ji}^{\pm}(\q) =& S_{ji}^{(x)} (\q) \pm i S_{ji}^{(y)}(\q)\, ,
\end{align}
where $S_{i,j}^{(\alpha)} ({\q})= (1/2)\sum_{{\p} s s'} \beta_{i{\p}s}^{\dagger} \sigma_{s,s'}^{(\alpha)}\beta_{j{\p+\q}s'}$,
and $\sigma^{(\alpha)}$ with $\alpha = x,y,z$ are Pauli matrices.
In Eq.~\eqref{generalized} $\q',\q''=\q,\q+\Q$.
The hybridization in 1FeBZ formulation is manifested
in the off-diagonal (umklapp) susceptibilities with $\q'-\q'' = \pm \Q$.
The $8$ by $8$ susceptibility matrix \cite{Brydon2009,Knolle2011} reads
\begin{align}\label{chi-matrix}
{\hat \chi} = \begin{bmatrix}
\hat{\chi} (\q,\q)   &  \hat{\chi} (\q,\q+\Q)   \\
\hat{\chi} (\q+\Q,\q) & \hat{\chi} (\q+\Q,\q+\Q)
\end{bmatrix} .
\end{align}

\begin{figure}
\begin{center}
\includegraphics[width=1.0\columnwidth]{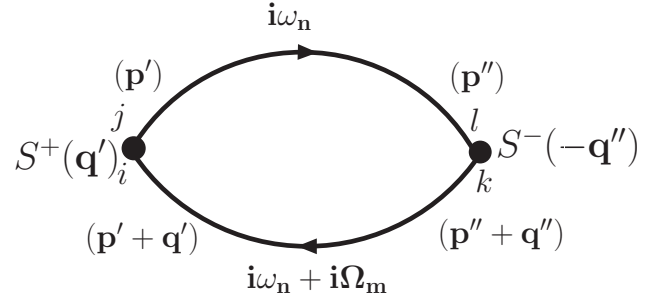}
\caption{Diagrammatic representation of
the contribution to  $\chi^0_{ijkl} (\q',\q'')$ from two normal Green's functions $G$.
The fermion momenta $\p$ and $\p'$ are either identical or differ by $\Q$.
The contribution from the anomalous Green's function $F$ has the same form, but single arrowed lines are replaced by the double arrowed lines representing anomalous propagators.
 }
\label{diagram}
\end{center}
\end{figure}

With band indices labeled as $1=\beta_1$ and $2=\beta_2$ each of the four susceptibility matrices in Eq.~\eqref{chi-matrix} has the following structure,

\begin{equation}\label{hat_chi_A}
\hat{\chi}(\q',\q'')=\!
\bordermatrix{
  & 11 & 22 & 12 & 21 \cr
11 & \chi_{1111} & \chi_{1122} & \chi_{1112} & \chi_{1121} \cr
22 & \chi_{2211} & \chi_{2222} & \chi_{2212} & \chi_{2221} \cr
12 & \chi_{1211} & \chi_{1222} & \chi_{1212} & \chi_{1221} \cr
21 & \chi_{2111} & \chi_{2122} & \chi_{2112} & \chi_{2121} \cr
}\, ,
\end{equation}
where the momenta arguments $(\q',\q'')$ were omitted on a right hand side for clarity.
Each entry in Eq.~\eqref{hat_chi_A} is defined by Eq.~\eqref{generalized}.
The dynamical spin structure factor, $S(\q,\omega)$ is obtained by summing over the entries of a matrix Eq.~\eqref{chi-matrix},
\begin{align}\label{physical}
S(\q,\omega)
 \propto
 \sum_{ijkl} \mathrm{Im}\left[\chi_{ijkl} (\q,\q)\right]\, .
\end{align}

We follow earlier works on the spin resonance in unconventional superconductors~\cite{Eschrig2006} and compute $S(\q,\omega)$ in the random phase approximation (RPA).
We have
\begin{eqnarray}\label{RPA}
{\hat \chi}= (\hat {1} - \hat{\chi}^{0} \hat{\Gamma})^{-1} \hat{\chi}^{0}.
\end{eqnarray}
In Eq.~\eqref{RPA}, $\hat{\chi}^{0}$ is the $8$ by $8$ bare susceptibility with the entries $\chi_{ijkl}^{0} (\q',\q'')$ shown schematically in Fig.~\ref{diagram}.
We express these matrix elements in terms of normal and anomalous Green's functions, Eq.~\eqref{Green-BCS-def}, in Appendix~\ref{sec:app_1}.
The interaction amplitude ${\hat \Gamma(\q,\q')} = \delta_{\q,\q'} \Gamma_{ijkl}$ follows from Eq.~\eqref{eq:2}.
The non-zero matrix elements are $\Gamma_{ijkl} = u_1,u_2,u_3,u_4$ for $i=k \ne j=l$, $i=j \ne k=l$, $i=l \ne k=j$, $i=j=k=l$, respectively.
In the numerical analysis of the resonance we used $u_1=u_3=1.95$ eV and $u_2=u_4=0.1$ eV. We verified that for these parameters, the normal state remains paramagnetic.

\subsection{The $s^{+-}$ ordered state}
\label{sec:The_s}
The quadratic Hamiltonian, $H_2 + H_{hyb}$, \eqref{eqH}, \eqref{Hhybrid} can be diagonalized~\cite{Khodas2012a} by transforming it to the basis of bonding and anti-bonding states ($ab$ basis),
\begin{align}\label{ab_basis}
a_{\p} & = \beta_{1\p} \cos \theta_{\p} + \beta_{2\p+\Q}\sin \theta_{\p}\, ,
\notag \\
b_{\p} & = - \beta_{1\p} \sin \theta_{\p}  +\beta_{2\p+\Q} \cos \theta_{\p}
\end{align}
where
\begin{align}\label{sin}
\sin 2 \theta_{\p} & = \frac{\lambda}{ \sqrt{ \lambda^2 + ( \delta \varepsilon_{\p})^2/4 }}\, ,
\notag \\
\cos 2 \theta_{\p} & = \frac{\delta \varepsilon_{\p} /2}
{ \sqrt{ \lambda^2 + ( \delta \varepsilon_{\p})^2/4 }}\, ,
\end{align}
and
\begin{align}\label{deltaEp}
\delta \varepsilon_{\p} =  \vare^{\beta_1}_{\p} - \vare^{\beta_2}_{\p+\Q}
\end{align}

In the $s^{+-}$-symmetric state the SC gap changes sign between the hybridized Fermi pockets. The pairing Hamiltonian reads
\begin{align}\label{H_s}
 H_s =& \Delta \sum_{\p} ( a_{\p} a_{-\p} - b_{\p} b_{-\p}) + h.c.  \nonumber \\
=& \Delta \sum_{\p} \left [ \cos 2\theta_{\p} ( \beta_{1\p} \beta_{1-\p} - \beta_{2\p+\Q} \beta_{2-\p-\Q} )
\nonumber \right. \\
+&
\left.
\sin 2\theta_{\p} ( \beta_{1\p} \beta_{2-\p-\Q}  + \beta_{2\p+\Q} \beta_{1-\p} )
\right ] + h.c.
\end{align}
In principle, $\Delta$ can have angle dependence, consistent with $s-$wave symmetry, but this dependence is not essential for our purposes and we neglect it.

To verify that the gap function defined by Eq.~\eqref{H_s} is $s$-wave symmetric we consider how it transforms under the rotation $\p \rightarrow \p'=(p_y,-p_x,p_z)$, $\beta_{1\p} \rightarrow \beta_{2\p'}$.
The invariance of Eq.~\eqref{H_s} follows from the properties $\cos 2 \theta_{\p'+\Q} = - \cos 2 \theta_{\p}$ and $\sin 2 \theta_{\p'+\Q} = \sin 2 \theta_{\p}$ easily derivable from Eqs.~\eqref{sin}, \eqref{deltaEp} and the dispersion relation Eq.~\eqref{dispersion}.
The gap parameters entering Eqs.~\eqref{H-BCS} can be read off the Eq.~\eqref{H_s} using Eq.~\eqref{sin} and have the form
\begin{subequations}
\begin{alignat}{1}
\Delta^{\beta_1,\beta_1}_{\p} =
-\Delta^{{\bar \beta_2},{\bar \beta_2}}_{\p}=
\Delta \frac{\delta \varepsilon_{\p}/2}{\sqrt{ \lambda^2 + (\delta\varepsilon_{\p})^2/4 }}\, ,\\
\Delta^{\beta_1,{\bar\beta_2}}_{\p} = \Delta^{{\bar\beta_2},\beta_1}_{\p} =
\Delta \frac{\lambda}{\sqrt{ \lambda^2 + (\delta\varepsilon_{\p})^2/4 }}\, .
\end{alignat}
\label{Eq-Gap}
\end{subequations}
\noindent
 Equations \eqref{dispersion} and \eqref{Eq-Gap} specify the mean field Hamiltonian \eqref{H-BCS}.

\section{Spin resonance in an $s^{+-}$ superconductor}
\label{sec_results}

\subsection{Spin resonance in $s^{+-}$ state at $\lambda=0$}
\label{sec:No_lambda}

As a warm-up, consider first the case when the  hybridization  is zero, i.e., $\lambda = 0$.
This limit is artificial because the $s^{+-}$ pairing is driven by hybridization and therefore requires a finite $\lambda$. Nevertheless,
 it is instructive to understand how the resonance develops at $\lambda =0$ before considering the actual case of a finite $\lambda$.
At $\lambda =0$,  the Cooper pairs are formed by electrons from the same band and have a zero center of mass momentum (the term with $\sin 2 \theta_p$ in Eq.\eqref{H_s} vanishes).
Correspondingly the OP Eq.~\eqref{Eq-Gap} is purely intra-band,
\begin{align}\label{Delta11}
\Delta^{11}_{\p} = \Delta \mathrm{sgn}(\delta_{\varepsilon_\p})\, , \,\,\, \Delta^{22}_{\p} =- \Delta \mathrm{sgn}(\delta_{\varepsilon_{\p+\Q}})\, .
\end{align}
To analyze the resonance, we then need to understand what happens when we connect parts of the {\it same} Fermi surface connected by ${\Q} = (\pi,\pi,\pi)$.
Eqs.~\eqref{Delta11} and \eqref{deltaEp} indicate that the OP changes sign across the lines defined by the condition $\varepsilon_{1\p} = \varepsilon_{2\p+\Q}$, i.e. along the lines of crossing of one Fermi pocket with the other shifted by $\Q$.
We recall that the sign changing of the OP is the necessary condition for spin resonance

\begin{figure}
\begin{center}
\includegraphics[width=1.0\columnwidth]{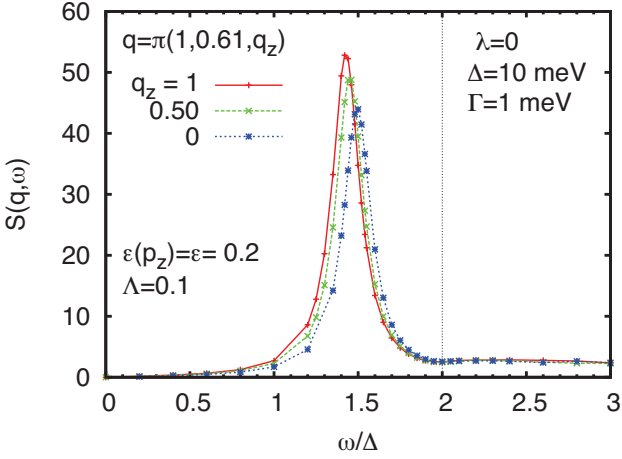}
\end{center}
\caption{Frequency dependence of the
spin structure factor
  $S(\q,\omega)$ for the {\it weak} dispersion limit. We set
   $\epsilon = 0.2$.
The wave-vector is $\q =(\pi, 0.61\pi,q_z)$ and
the values of $q_z$ for three different curves are  $q_z =\pi$, $q_z = 0.5 \pi$, and $q_z = 0$.
The dispersion parameter is set at
$\Lambda =0.1$ and the gap $\Delta = 10$meV.
A small imaginary component, $\Gamma=1$meV, is added to frequency for regularization.
}
\label{figL_NO_lambda}
\end{figure}

In the weak (strong) dispersion limit the lines across which the OP changes sign are approximately vertical (horizontal), see Fig.~\ref{f2}.
In the weak dispersion limit, the origin of the spin resonance in our case is qualitatively similar to that in the situation when superconducting gap has a  $d$-wave symmetry~\cite{Maier2011a}. Our results for this case are presented in  Fig.~\ref{figL_NO_lambda}.
In the case of  strong dispersion, there are new pieces of physics which are worth discussing before moving to the case $\lambda \neq 0$.

\begin{figure}
\begin{center}
\includegraphics[width=1.0\columnwidth]{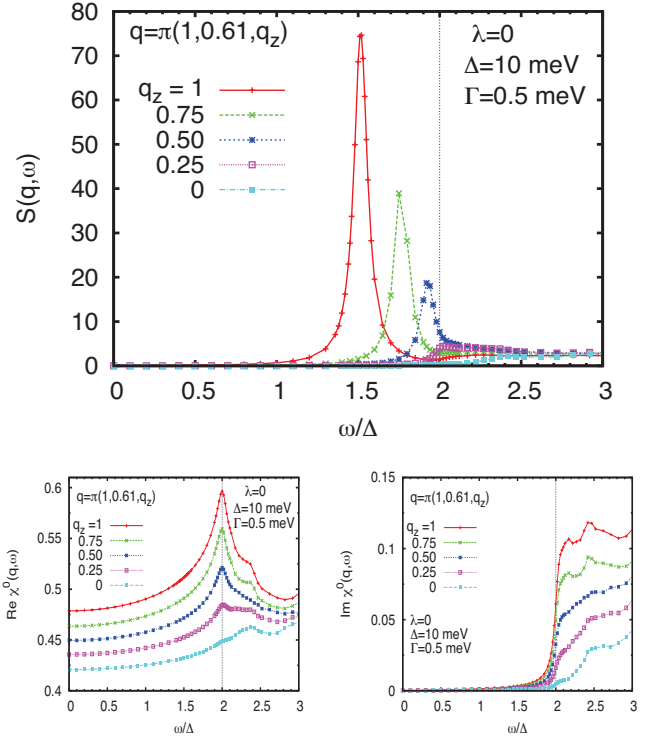}
\end{center}
\caption{
Upper panel. Frequency dependence of the
 spin structure factor
   $S(\q,\omega)$ for the {\it strong} dispersion limit. Now
    the ellipticity, $\epsilon(p_z) = \epsilon \cos(p_z)$ with $\epsilon=0.1$, 
changes sign at $p_z =\pm\pi/2$
The
the values of $q_z$ for five different curves are
 $\q =(\pi, 0.61\pi,q_z)$ and $q_z =\pi$, $q_z = 0.75 \pi$, $q_z = 0.5 \pi$, $q_z = 0.25 \pi$ and $q_z = 0$.
The dispersion parameter $\Lambda =0.1$ and the gap $\Delta = 10$meV.
A small imaginary component, $\Gamma=0.5$meV, is added to the frequency for regularization.
Lower panel. Frequency dependence of the real and imaginary part of $\chi_{12}^0 + \chi_{21}^0$ (see Eq.~\eqref{bare_susc}),
 shown for the same set of parameters as for the upper panel.
}
\label{fig-SC-Chi0-RPA-Lam0}
\end{figure}

\begin{figure}
\includegraphics[width=0.9\columnwidth]{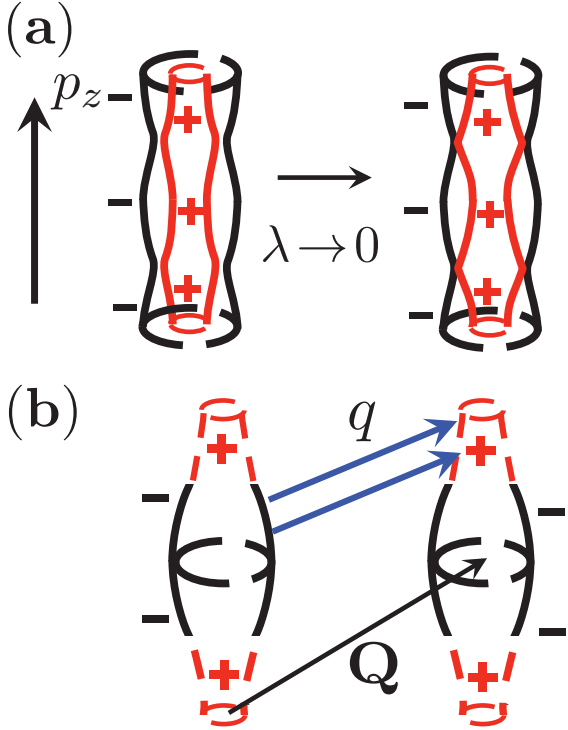}
\caption{(color online)
 The  limit $\lambda =0$.
(a) Superconducting gap for
 $s^{+-}$ pairing symmetry.
(b) The unfolded Fermi pockets.
The gap changes sign at
$p_z = \pm \pi/2$.
At these momenta the two
 Fermi pockets cross in the folded BZ.
The folding vector $\Q$ is shown in black.
For a
 given
  wave-vector $\q$,
 only states on a  portion of the Fermi surface contribute to the resonance (points connected
  by thick (blue) arrowed lines).
For $q_z = \pi$ all states on the Fermi surface
 are involved.
 For $q_z=0$
 the transitions are horizontal. In this limit, the transitions only occur between states on a Fermi surface with the same sign of the gap
  and the resonance does not develop.
}
\label{fig-FS-folded}
\end{figure}

Our  numerical results for this case are shown in the upper panel of Fig.~\ref{fig-SC-Chi0-RPA-Lam0}.
We see that the resonance weakens progressively as $q_z$ decreases from $\pi$ to $0$.
To understand this, we notice that the OP on each of unhybridized Fermi surfaces changes sign across the horizontal planes, $|p_z| = \pi/2$, see Fig.~\ref{fig-FS-folded}.
As a result, at $q_z = \pi$ the gaps on all points of the two pieces of the {\it same} Fermi surface connected by ${\Q} = (\pi,\pi,\pi)$ have opposite sign.
In contrast, $q_z = 0$ connects  Fermi surface points with the same sign of the superconducting OP.
Outside of the limit of strong $p_z$ dependence, the OP changes the sign along a line not necessarily confined to a constant $p_z$ plane, and the resonance in general is expected at all $q_z$ as is indeed the case for weak $p_z$ dispersion (Fig~\ref{figL_NO_lambda}).


To justify this argumentation, we analyze below a general expression for the spin susceptibility.
In the absence of the hybridization, the umklapp susceptibility in Eq.~\eqref{chi-matrix} vanishes and the bare spin susceptibility matrix $\hat{\chi}(\q,\q')$ in Eq.~\eqref{hat_chi_A} becomes diagonal
\begin{align}\label{bare_susc}
\chi_{ijkl}^0 (\q,\q';\omega) = \delta_{\q,\q'} \delta_{ik} \delta_{lj} \chi_{ij}^0 (\q,\omega)\, .
\end{align}
In this case $\chi_{ij}^0 (\q,\omega)$ can be expressed explicitly as
\begin{align}\label{chi^0}
\chi^0_{ij}  (\q,\omega) = & \frac {1}{4}\sum_{\p} \Bigg[
C^{(1)}_{ij;\p,\q}
 \frac {f(E_{\p+\q}^j) - f(E_{\p}^i)} {\omega + i 0^+ -(E_{\p+\q}^j - E_{\p}^i)}
 \notag \\
&+
C^{(2)}_{ij;\p,\q}  \frac {f(E_{\p}^i) -
 f(E_{\p+\q}^j)} {\omega + i 0^+ -(E_{\p}^i - E_{\p+\q}^j)}
\notag \\
&+
C^{(3)}_{ij;\p,\q}  \frac {1-f(E_{\p}^i)  - f(E_{\p+\q}^j)} {\omega + i 0^+ +(E_{\p}^i + E_{\p+\q}^j)}
\notag \\
&
+
C^{(4)}_{ij;\p,\q}
 \frac {f(E_{\p}^i) +
 f(E_{\p+\q}^j)-1} {\omega + i 0^+ -(E_{\p}^i + E_{\p+\q}^j)}
 \Bigg],
\end{align}
where the $f(E)$ is Fermi distribution function and coherence factors are
\begin{align}\label{Coherence}
C^{(1)}_{ij;\p,\q} & = 1+\frac{\vare_{\p}^i}{E_{\p}^i} + \frac{\vare_{\p+\q}^j}{E_{\p+\q}^j} +
 \frac{\vare_{\p}^i \vare_{\p+\q}^j + \Delta_{\p}^{i}
 \Delta_{\p+\q}^{j}} {E_{\p}^i E_{\p+\q}^j}
\notag \\
C^{(2)}_{ij;\p,\q} & = 1-\frac{\vare_{\p}^i} {E_{\p}^i} - \frac{\vare_{\p+\q}^j}{E_{\p+\q}^j} +
 \frac{\vare_{\p}^i \vare_{\p+\q}^j + \Delta_{\p}^i
 \Delta_{\p+\q}^j} {E_{\p}^i E_{\p+\q}^j}
 \notag \\
 C^{(3)}_{ij;\p,\q} & = 1+\frac{\vare_{\p}^i}{E_{\p}^i} - \frac{\vare_{\p+\q}^j}{E_{\p+\q}^j} -
 \frac{\vare_{\p}^i \vare_{\p+\q}^j + \Delta_{\p}^i
 \Delta_{\p+\q}^j} {E_{\k}^i E_{\p+\q}^j}
\notag \\
 C^{(4)}_{ij;\p,\q} & =1-\frac{\vare_{\p}^i} {E_{\p}^i} + \frac{\vare_{\p+\q}^j} {E_{\p+\q}^j} -
 \frac{\vare_{\p}^i \vare_{\p+\q}^j + \Delta_{\p}^i
 \Delta_{\p+\q}^j} {E_{\p}^i E_{\p+\q}^j}.
 \end{align}
 The  mean field quasi-particle energy is
\begin{align}\label{Ep}
E_{\p}^{1(2)} = \sqrt{(\varepsilon^{1(2)}_{\p})^2+(\Delta_{\p}^{1(2)})^{2}}\, .
\end{align}
In Eqs.~\eqref{Coherence}, \eqref{Ep} and below we set $\Delta^{ii} \equiv \Delta^{i}$, $i=1,2$.

At low temperatures the last (fourth) term in Eq.~\eqref{chi^0} makes a dominant contribution to $S(\q,\omega)$.
The intra-band susceptibilities ($i =j$ in Eq.~\eqref{chi^0}) are much smaller than the inter-band ones ($i \neq j$ in Eq.~\eqref{chi^0}) at the momenta ${\bf q} \approx (\pi,\pi)$.
Indeed, the energy of an intra-band excitations at such momentum is of the order of the bandwidth,
which is much larger than the typical energy of inter-band excitations at the same momentum.
As a result, the susceptibilities $\chi_{ii}^{0}$ are suppressed by the large energy denominators.
We have verified numerically  that the band diagonal susceptibilities do not affect the spin structure factor.
In this situation, the in-gap spin collective mode is due to the singularity of inter-band susceptibilities at the threshold of the particle-hole continuum ($\omega = 2 \Delta$).
The stronger the singularity the more pronounced is the spin resonance, as it is clearly seen in Fig.~\ref{fig-SC-Chi0-RPA-Lam0}.
The inter-band susceptibility $\chi^0_{12}$ is singular provided the coherence factors $C^{(3,4)}_{12;\p,\q}$ in Eq.~\eqref{Coherence}  do not vanish at the Fermi surface,
($\vare_{\p}^1,\vare_{\p+\q}^{2}\rightarrow 0$), i.e. provided that
$\left(1- \frac {\Delta_{\p}^1 \Delta_{\p+\q}^2} {\left\vert \Delta_{\p}^1 \right\vert \left\vert \Delta_{\p+\q}^2 \right\vert } \right) \neq 0$.
To put it simply, the resonance appears for neutron momentum $\q$ connecting regions of the two Fermi pockets with different sign of $\Delta$.
At $q_z =0$ the susceptibility becomes regular, and the resonance disappears.
We will see in the next section that at finite $\lambda$, $\chi^0$ retains the singularity even at $q_z =0$.

\begin{figure}
\includegraphics[width=1.0\columnwidth]{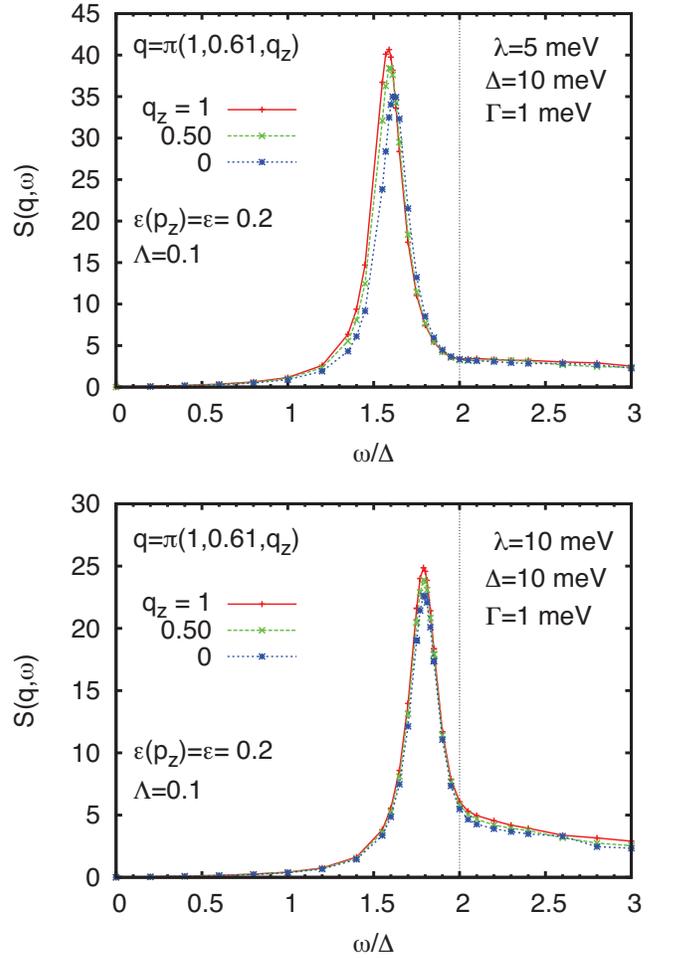}
\caption{Frequency dependence of the spin structure factor
  $S(\q,\omega)$ in the {\it weak} dispersion limit ($\epsilon = 0.1$) at a finite hybridization $\lambda = 5$meV (top) and $\lambda = 10$meV (bottom).
The wave-vectors for three different curves are $\q =(\pi, 0.61\pi,q_z)$ and $q_z =\pi$, $q_z = 0.5 \pi$, and $q_z = 0$.
The dispersion parameter $\Lambda =0.1$ and the gap $\Delta = 10$meV.
The small imaginary component, $\Gamma=1$meV is added to frequency for regularization.}
\label{weak_lambda}
\end{figure}

\begin{figure}
\includegraphics[width=1.0\columnwidth]{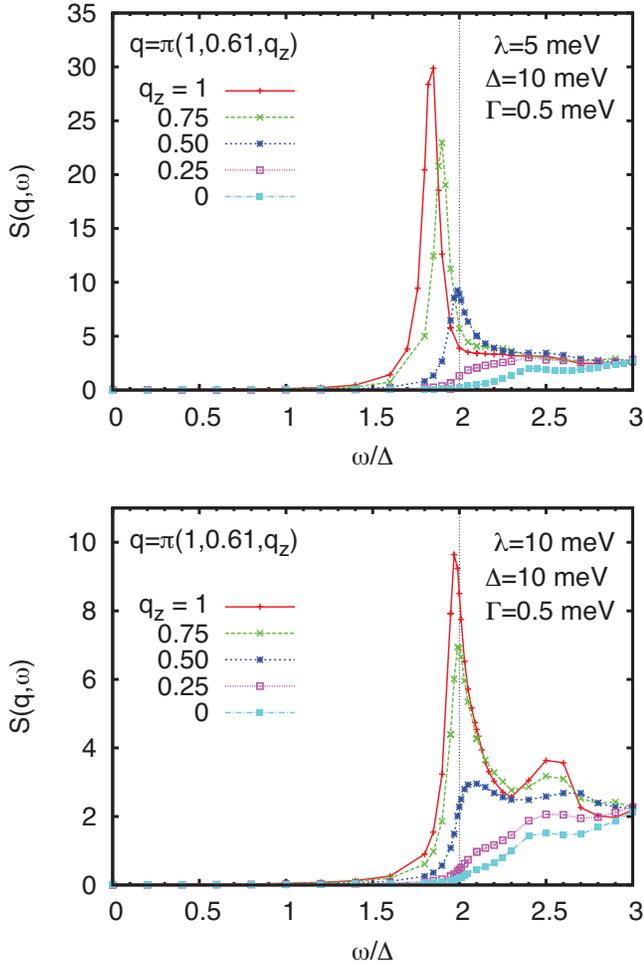}
\caption{The spin structure factor $S(\q,\omega)$ in the {\it strong} dispersion limit ($\epsilon = 0.1\cos(p_z)$) at a finite hybridization $\lambda=5$meV (upper panel) and for $\lambda=10$meV (lower panel).
The curves are plotted for five wave-vectors, $\q =\pi(1,0.61,q_z)$, where $q_z = 1, 0.75, 0.5, 0.25, 0$.
The resonance is present for all $q_z$, but it gets weaker with decreasing $q_z$ and with increasing hybridization.}
\label{fig-SC-qz-lam}
\end{figure}

\subsection{Spin resonance in $s^{+-}$ superconductor at a finite $\lambda$}
\label{sec:finite}

As in the previous section, we discuss separately the cases of the weak and strong band dispersion.

The results for the weak dispersion limit are shown in Fig.~\ref{weak_lambda}.
We see that with increasing hybridization the spin resonance weakens and becomes more two-dimensional.  This result is entirely expected.

The effect of the hybridization on the spin resonance in the strong dispersion limit is more nuanced.
Our numerical results for the spin structure factor in this limit and at a finite hybridization are presented in Fig.~\ref{fig-SC-qz-lam}.
The key result is that the
resonance is clearly seen for large subset  of $q_z$ values except for a small range near $q_z =0$.
Below we argue that the suppression of the resonance near $q_z =0$ is non-generic, and for a generic dispersion relation the resonance is expected to be present for all $q_z$s.

To understand the influence of the hybridization on the resonance it is useful to consider the spin operator in the basis of bonding and anti-bonding states ($a$ and $b$ states in Eq.~\eqref{ab_basis}).
The singular part of the spin susceptibility is determined by the coherence factor and by the matrix element of the spin operator connecting bonding and anti-bonding states.
In $ab$-basis the coherence factor is a constant (see Eq. \eqref{H_s}).
The matrix element is obtained by writing the inter-band spin operator,
$S^{+}_{\mathrm{eff}}(\q) = S^+_{12}(\q) + S^+_{21}(\q)$, defined by Eq.~\eqref{generalized}, in terms of $a_{\p}$ and $b_{\p}$ operators using Eq.~\eqref{ab_basis}.
Keeping only the off-diagonal ($ab$) components we obtain
\begin{align}\label{S_eff}
S^{+}_{\mathrm{eff}}(\q) \approx & \sum_{\p} M_{\p,\delta \q}
\left( a_{\p \uparrow}^{\dag} b_{\p +\delta \q \downarrow} + b_{\p \uparrow}^{\dag} a_{\p+\delta \q \downarrow} \right)\, ,
\notag \\
M_{\p,\delta \q} = &\left( \cos \theta_{\p} \cos \theta_{\p + \delta \q} -  \sin \theta_{\p} \sin \theta_{\p + \delta \q} \right)\, ,
\end{align}
where we represent the scattering momentum, $\q$ in the form $\q = \Q + \delta \q$, such that the vector $\delta q = \delta q_x \hat{x} + \delta q_y \hat{y} + \delta q_z \hat{z}$ has  small $xy$ components, $\delta q_x,\delta q_y \ll \pi$.
The strength of the resonance is determined by the matrix element for an inter-band transition with the spin flip, as given by Eq.~\eqref{S_eff}.
For the transition probability we evaluate the squared matrix element using Eqs.~\eqref{sin} and \eqref{deltaEp}.
We obtain
\begin{align}\label{Mp}
|M_{\p,\delta \q}|^2 = & \frac{1}{2} + \frac{ 1}{2}
\frac{ \delta \varepsilon_{\p} \delta \varepsilon_{\p + \delta \q}  -4 \lambda^2 }{ \sqrt{ 4 \lambda^2 + (\delta \varepsilon_{\p})^2 }  \sqrt{4 \lambda^2 + (\delta \varepsilon_{\p+\delta \q})^2 } }   \, .
\end{align}
We argue, based on Eq.~\eqref{Mp}, that generally the resonance is the strongest at $q_z = \pi$ ($\delta q_z = 0$),
as it was the case without hybridization.
However, the hybridization affects the resonance at $q_z = \pi$ ($\delta q_z = 0$) and  $q_z = 0$ ($\delta q_z = \pi$) in an opposite way -- it suppresses the resonance at $q_z = \pi$ and makes it non-zero at $q_z = 0$.
 This trend persists as long as $\lambda$ does not exceed a certain magnitude $\lambda \lesssim |\delta \varepsilon_{\p}|$.
With further increase of hybridization, the resonance is suppressed for all $q_z$ because matrix element $M_{\p,\delta \q}$ for $\lambda \gg \delta\varepsilon_{\p}$ gets smaller, see Eq.~\eqref{Mp}.

\begin{figure}
\includegraphics[width=1.0\columnwidth]{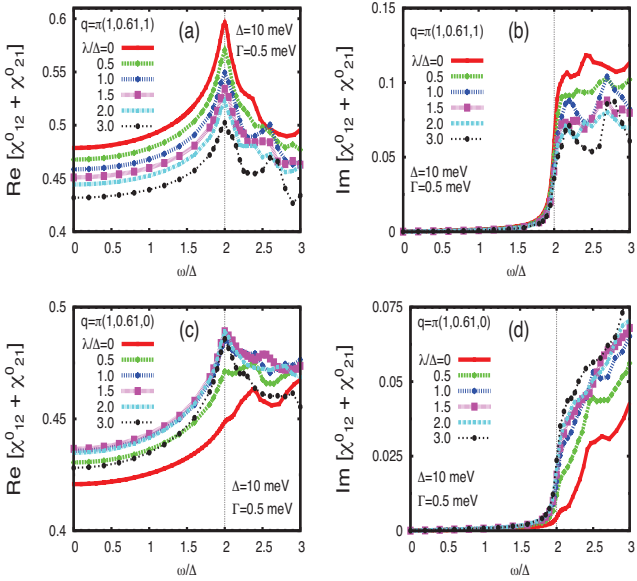}
\caption{Effect of the hybridization on the singularity in the bare spin susceptibility at $\omega = 2 \Delta$ (the peak in  $\mathrm{Re}[\chi^0_{12} + \chi^0_{21}]$
 and the  jump in $\mathrm{Im}[\chi^0_{12} + \chi^0_{21}]$, where $\chi^0_{12} \equiv \chi^0_{1212}$ and $\chi^0_{21} \equiv \chi^0_{2121}$).
Panels (a) and (b) -- real and imaginary parts of $\chi^0_{12} + \chi^0_{21}$ for
$\q = \pi( 1,0.61,1)$. Both the peak and the jump are
 suppressed when $\lambda$ increases.
Panels (c) and (d)  -- the same for $\q = \pi( 1,0.61,0)$. The trends with increasing $\lambda$ is the opposite -- both the peak and the jump get larger.
The other inter-band susceptibilities, $\chi^0_{1221}$ and $\chi^0_{2112}$, contribute much less to the singularity in the susceptibility,
 these contributions are negative and
 increase with hybridization independent of the value of $q_z$.}
\label{indicate}
\end{figure}

The opposite effect of the hybridization on the intensity of the resonance at $q_z =0$ and $q_z =\pi$
are  clearly seen in our numerical calculations, see Fig.~\ref{indicate}.
For $q_z = \pi$ the characteristic peak (jump) in real (imaginary) part of the bare inter-band susceptibility is suppressed by hybridization, thereby making the resonance weaker,  Fig.~\ref{indicate}(a), (b).
For $q_z=0$, spin susceptibility becomes singular at a finite hybridization, see
Fig.~\ref{indicate}(c), (d),
which indicates that hybridization induces spin resonance at this $q_z$.
When the hybridization is increased further, the initial enhancement is reversed, and the spin resonance gets suppressed for all $q_z$.

To explain this non-monotonic $q_z$ dependence of the intensity of the resonance
we analyze the formula for $|M_{\p,\delta \q }|^2$, Eq.~\eqref{Mp}.
For $\q = \Q$, i.e $\delta \q = 0$,
\begin{align}\label{Mp1}
|M_{\p,\delta \q =0}|^2 =
\frac{ (\delta \varepsilon_{\p} )^2}{4 \lambda^2 + (\delta \varepsilon_{\p})^2 }   \, ,
\end{align}
reaches  the maximal value of $1$ at $\lambda = 0$ and is suppressed for non-zero $\lambda$.
This obviously implies that the resonance intensity gradually decreases when $\lambda$ increases.

Consider next $\q = (\pi,\pi,0)$, i.e $\delta \q = \pi \hat{z}$.
We have
\begin{align}\label{Mp2}
|M_{\p,\delta \q = \pi \hat{z}}|^2 = \frac{1}{2}
+ \frac{1}{2} \frac{\delta \varepsilon_{\p} }{ | \delta \varepsilon_{\p} | } \frac{\delta \varepsilon_{\p + \pi \hat{z}} }{ | \delta \varepsilon_{\p + \pi \hat{z}} | }
\end{align}
for $\lambda =0$ and
\begin{align}\label{Mp3}
|M_{\p,\delta \q = \pi \hat{z}}|^2 = \frac{1}{2}
+ \frac{1}{2} \frac{\delta \varepsilon_{\p} \delta \varepsilon_{\p + \pi \hat{z}} - 4 \lambda^2 }
{ \sqrt{4 \lambda^2 + (\delta \varepsilon_{\p})^2 }\sqrt{4 \lambda^2 + (\delta \varepsilon_{\p+ \pi \hat{z}})^2 } }
\end{align}
 for $\lambda \neq 0$.
The energy difference $\delta\varepsilon_{\p}$, Eq.~\eqref{deltaEp} changes sign at $p_z = \pi/2$ and $p_z =- \pi/2$, which are separated by momentum $\pi$ along $p_z$ direction.
Then $\mathrm{sgn}( \delta \varepsilon_{\p} ) = - \mathrm{sgn}( \delta \varepsilon_{\p + \pi \hat{z}} )$, and the matrix element in Eq.~\eqref{Mp2} vanishes.
This explains why there is no resonance at $q_z =0$
in the absence of hybridization.
The same argument also makes it clear that the resonance is expected for more generic band structure with $\delta \epsilon_{\p}$ vanishing along arbitrary line not confined to a constant $p_z$.
At a finite $\lambda$, the matrix element Eq.~\eqref{Mp3} vanishes if and only if the condition
\begin{align}\label{condition}
\delta \varepsilon_{\p} + \delta \varepsilon_{\p + \pi \hat{z} } = 0
\end{align}
is satisfied.
One can readily check that this condition does not hold for a general $\p$.
The sum in Eq.~\eqref{condition}, evaluated using Eqs.~\eqref{dispersion} and \eqref{deltaEp}, reduces to
\begin{align}
\delta \varepsilon_{\p} + \delta \varepsilon_{\p + \pi \hat{z} }
= 4 t \Lambda \epsilon(p_z) \cos p_z ( \cos p_x + \cos p_y )
\end{align}
is in general non-zero, although it is small when $\Lambda$ and $\epsilon$ are small.

Furthermore, we show in Appendix~\ref{app:B} that for $q_z = 0$,
$\mathrm{Re} \chi^0$ has a logarithmic singularity
at $\omega = 2 \Delta$.
 This singularity is obtained for $q_x,q_y$ such that one of the Fermi surfaces shifted by $(q_x,q_y,q_z)$ touches the other Fermi surface
 for all $q_z$. However, because the singularity is reduced by the smallness of the matrix element (when $\Lambda$ and $\epsilon (p_z)$ are small),
the binding energy of the resonance is small, and in practice the resonance can be washed out by lifetime effects.
In other words, the spin resonance does exist at all $q_z$ when hybridization is non-zero, but its intensity is the smallest at $q_z =0$.

\section{Conclusions}
\label{sec:conclusions}
In this paper we have demonstrated that the observed spin resonance in the alkali-intercalated iron selenides is consistent with $s^{+-}$  superconductivity in which superconducting gap
  changes sign between the hybridized bonding and anti-bonding bands.
We found that the existence of the gaps with different signs  does not necessarily lead to the appearance of the spin resonance.
In particular, there is no resonance for the case when the Fermi surfaces before hybridization are circular cylinders because in this situation
all states are hybridized into bonding and anti-bonding states which are even or odd, respectively, with respect to interchange between
 fermionic pockets.  In  $s^{+-}$  state the gap changes sign between bonding and antibonding Fermi surfaces, however, spin operator is symmetric with respect to interchange
between pockets and does not have a non-zero matrix element between bonding and anti-bonding states.
However, for elliptical pockets the resonance does exist because the splitting into
bonding and anti-bonding states holds only for a fraction of fermions located near
the crossing lines in 3D space between  one pocket and the other one, translated by a folding vector $\Q$.
For other fermions,  hybridization is a weak effect, and the states on the Fermi surfaces with ``plus'' and ``minus'' gap are
 coupled by the spin operator.

We found that the resonance exists for both weak and strong dispersion of fermionic excitations along the $z$ axis perpendicular to Fe planes.
For weak dispersion, the resonance is essentially a 2D phenomenon, and its energy and intensity weakly depend
on $q_z$. For strong dispersion,
 the intensity of the resonance is the strongest at $q_z = \pi$ and the weakest at $q_z =0$, where for the dispersion which we used, it only exists
  due to a finite hybridization.  Still, for realistic hybridization the resonance  becomes quasi-two-dimensional, and the optimal wave vector in $xy$ plane (at which the intensity is the largest)
   is close to $(\pi,\pi/2)$, consistent with what was  reported experimentally \cite{Park2011,Friemel2012,Friemel2012a,Taylor2012,Wang2012}.

\begin{acknowledgements}
The authors are grateful to R. Fernandes, P.J. Hirschfeld, D. Inosov, W. Ku, A. Levchenko, T.A. Maier, I.I. Mazin, J. Schmalian, M.G. Vavilov  for valuable discussions.
M.K. acknowledges the support of University of Iowa.  A.V.C. is supported by the Office of Basic Energy Sciences  U.S. Department of Energy under  the grant
\#DE-FG02-ER46900.
\end{acknowledgements}

\begin{appendix}
\section{Calculation of the matrix elements of a bare susceptibility $\hat{\chi}^0(\q',\q'')$}
\label{sec:app_1}
Fig.~\ref{fig:chi0-qq} shows the diagrammatic representation of the different matrix elements of the bare susceptibility $\hat{\chi}^0(\q',\q'', i\Omega_m)$.
First we consider the diagrammatic contributions (a) for $\chi_{ijkl}^{0}(\q,\q;i\Omega_m)$ which can be expressed as:

\begin{widetext}
\begin{align}
\chi_{ijkl}^{0}(\q,\q;i\Omega_m)&= -\frac{1}{2\beta}\sum_{\p,\omega_n}
 \left[ \left( G_{ik}  (\p+\q, i\omega_n + i\Omega_m) G_{lj} (\p, i\omega_n)  +  G_{{\bar i}{\bar k}} (\p+\q, i\omega_n + i\Omega_m)  G_{{\bar l}{\bar j}}
 (\p, i\omega_n)
\right. \right.
\nonumber \\
&+
\left. \left.
G_{i{\bar k}} (\p+\q, i\omega_n + i\Omega_m)  G_{{\bar l}j} (\p, i\omega_n) + G_{{\bar i} k} (\p+\q, i\omega_n +i\Omega_m) G_{l{\bar j}} (\p, i\omega_n)  \right ) +
  (  G \leftrightarrow F) \right ] \nonumber\\
\end{align}

Here it can easily be noticed the identical contributions of the first and second terms, which correspond to the two upper diagrams in (a), and also that of the third and fourth
terms, which correspond to the two lower diagrams in (a). Therefore, the above expression can be rewritten as

\begin{figure}
\begin{center}
\includegraphics[width=1.0\columnwidth]{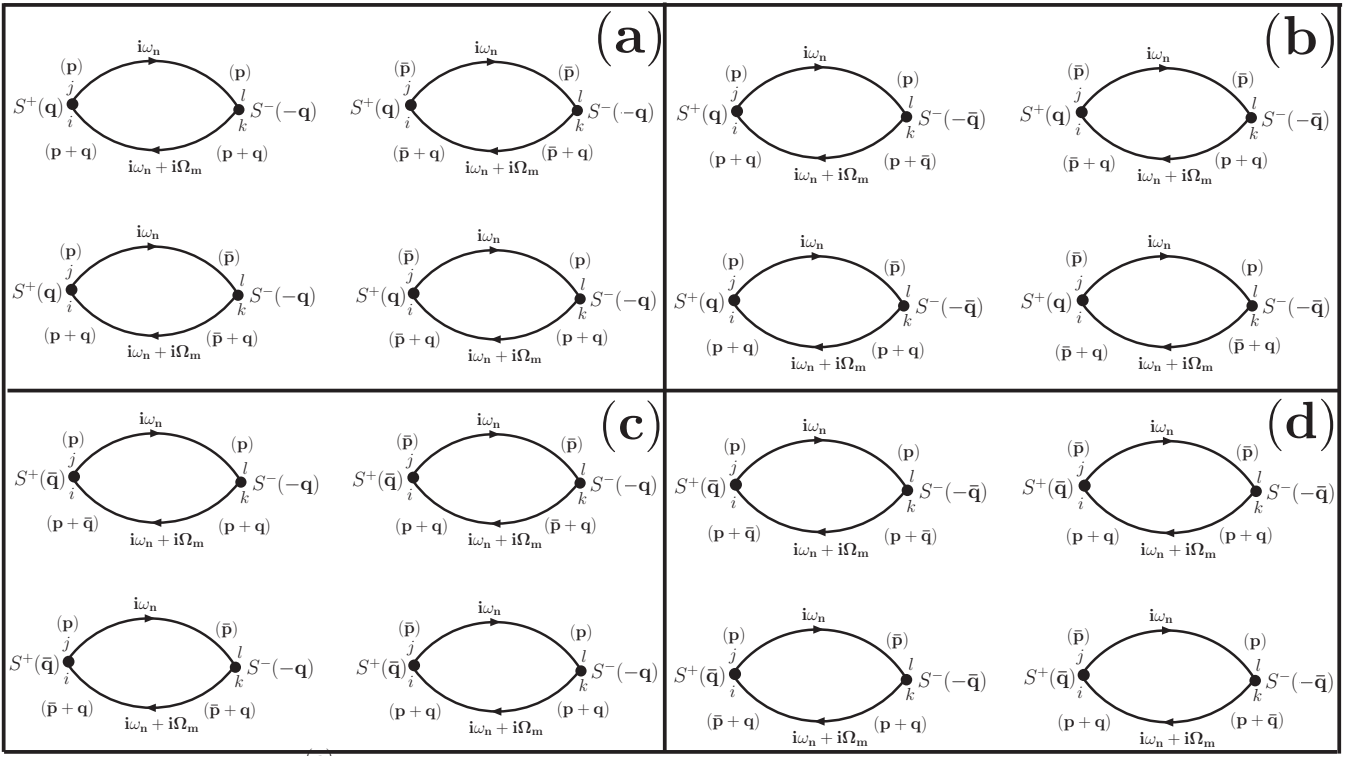}
\caption{
 Diagrammatic representation of the components of the bare particle-hole propagator.
 Each component of
$\chi^0_{ijkl} (\q',\q'';i\Omega_m)$ of both normal $(\q'=\q'')$ and umpklapp ($|\q'-\q''|= Q$) susceptibilities has four diagrammatic contributions.
The diagrams (a),(b),(c), and (d) represent the bare susceptibilities
$\chi^0_{ijkl} (\q,\q;i\Omega_m)$, $\chi^0_{ijkl} (\q,\bar\q;i\Omega_m)$, $\chi^0_{ijkl} (\bar\q,\q;i\Omega_m)$, and $\chi^0_{ijkl} (\bar\q,\bar\q;i\Omega_m)$, respectively where $\bar \q= \q+\Q$ (and  $\bar \p= \p+\Q$).
 Only the contributions from normal Green's functions $G$ are shown.
 The contributions from the anomalous Green's functions $F$  have the same form, but single arrowed lines are replaced by the double arrowed lines.
 representing anomalous propagators.
}
\label{fig:chi0-qq}
\end{center}
\end{figure}

\begin{align}
\chi_{ijkl}^{0}(\q,\q;i\Omega_m) &=
-\frac{1}{\beta}\sum_{\p,\omega_n}
 \left[ \left( G_{ik}  (\p+\q, i\omega_n + i\Omega_m) G_{lj} (\p, i\omega_n)  +
G_{i{\bar k}} (\p+\q, i\omega_n + i\Omega_m)  G_{{\bar l}j} (\p, i\omega_n)\right )
+ (  G \leftrightarrow F) \right ]
\label{eq:chi0-qq}
 \end{align}

Similarly, by taking into account the identical contributions of the two upper and the two lower diagrams also in (b), (c), and (d),  the expressions for
$\chi_{ijkl}^{0}(\q,\bar \q;i\Omega_m)$ (b), $\chi_{ijkl}^{0}(\bar \q,\q;i\Omega_m)$ (c), and
$\chi_{ijkl}^{0}(\bar \q, \bar \q;i\Omega_m)$ (d) can be written as:

\begin{align}
\chi_{ijkl}^{0}(\q,\bar \q;i\Omega_m)&= -\frac{1}{\beta}\sum_{\p,\omega_n}
 \left[ \left( G_{i{\bar k}}  (\p+\q, i\omega_n + i\Omega_m) G_{lj} (\p, i\omega_n) +
G_{ik} (\p+\q, i\omega_n + i\Omega_m) G_{{\bar l}j} (\p, i\omega_n) )   \right ) +
  (  G \leftrightarrow F) \right ]
 \label{eq:chi0-qqQ}
 \end{align}

\begin{align}
\chi_{ijkl}^{0}(\bar \q,\q;i\Omega_m)&= -\frac{1}{\beta}\sum_{\p,\omega_n}
\left[ \left(  G_{{\bar i} k} (\p+\q, i\omega_n + i\Omega_m) G_{lj} (\p, i\omega_n) +
 G_{ik} (\p+\q, i\omega_n + i\Omega_m) G_{l{\bar j}} (\p, i\omega_n) \right ) +  (  G \leftrightarrow F) \right ]
 \label{eq:chi0-qqQ1}
\end{align}

\begin{align}
\chi_{ijkl}^{0}(\bar \q, \bar \q;i\Omega_m)&=-\frac{1}{\beta}\sum_{\p,\omega_n}
 \left[ \left(G_{{\bar i}{\bar k}}(\p+\q, i\omega_n + i\Omega_m) G_{lj} (\p, i\omega_n)
+ G_{{\bar i} k} (\p+\q, i\omega_n + i\Omega_m) G_{{\bar l}j}(\p, i\omega_n) \right ) + (  G \leftrightarrow F) \right ].
 \label{eq:chi0-qqQQ}
 \end{align}

The matrix elements $\chi_{ijkl}^0$ can be written in terms of the normal and anomalous Green's functions by identifying the Green's functions entering
Eqs.~\eqref{eq:chi0-qq}, \eqref{eq:chi0-qqQ}, \eqref{eq:chi0-qqQ1} and \eqref{eq:chi0-qqQQ} with matrix elements of ${\cal {\hat G}}$ in Eq.~\eqref{Green-BCS-def}.
It is expedient to diagonalize it to
 \begin{equation}
{\hat {\cal G}}_{\alpha\beta} (p,i\omega_n)= \sum_{\mu} \frac{a_\mu^{\alpha}  a_\mu^{\beta*}}
{i\omega_n - E^\mu_{\p}} \, ,
\label{Eq:G-function}
\end{equation}
where the indices $\alpha$, $\beta$ run from 1 to 4.
The eigenvalues $E^{\mu}_{\p}$ and eigenvectors $a_\mu^{\alpha}$ are labeled by an index $\mu=1,2,3,4$.
The frequency summation can then be performed analytically.

For illustration, we evaluate the matrix element $\chi^0_{1111} (\q,\q,\Omega_m)$.
Since, in this particular case $G_{1{\bar 1}} = G_{{\bar 1}1} = F_{1{\bar 1}} =
F_{{\bar 1}1} =0$, the expression for $\chi^0_{1111}$ simplifies to

\begin{eqnarray}
\chi_{1111}^{0}(\q,\q;i\Omega_m)=
 -\frac{1}{\beta}\sum_{\p,\omega_n}
 \left[ G_{11}(\p,i\omega_n) G_{11} (\p', i\omega_n + i\Omega_m) + (  G \leftrightarrow F) \right]\, ,
  \label{eq:chi0-qq-1}
\end{eqnarray}
where $\p'=\p+\q$. Now identifying $G_{11}(\p)={\hat {\cal G}}_{11}$,  $F_{11}(\p)={\hat {\cal G}}_{12} (\p)$ in Eq.~\eqref{Green-BCS-def} and using Eq.~\eqref{Eq:G-function} the sum over fermion Matsubara frequencies $\omega_n$ can be carried out analytically which yields after analytic continuation $i \Omega_m \rightarrow \omega+i0$.

\begin{align}
\chi_{1111}^{0}(\q, \q; \omega)
= & \sum_{\p,\mu,\nu}
\left[
a_\mu^{1}(\p) a_\mu^{1*} (\p) a_\nu^{1}(\p+\q) a_\nu^{1*} (\p+\q) +
a_\mu^{1}(\p) a_\mu^{2*} (\p) a_\nu^{1}(\p+\q) a_\nu^{2*} (\p+\q) \right]
\notag \\
& \times
\left[ \frac{f(E^\nu_{\p+\q}) - f(E^\mu_{\p})}{\omega + i 0^+ - (E^\nu_{\p+\q}  -E^\mu_{\p}) } \right] \, ,
\label{eq:chi0_11}
\end{align}
where $f(E)$ is the Fermi function.
In numerical calculations a small imaginary part $\Gamma$ is added to the frequency $\omega$ for regularization, $\omega \rightarrow \omega + i \Gamma$.
\end{widetext}
\section{Threshold singularities of $\chi^0$ at $\omega \rightarrow 2 \Delta$}
\label{app:B}
At low temperatures the bare susceptibilities $\chi^0$ is determined by the last term of Eq.~\eqref{chi^0},
\begin{align}\label{app:chi^0}
\chi^0_{ij}  (\q,\omega) =  \frac {1}{4}\sum_{\p}
C^{(pp)}_{ij;\p,\q}
 \frac {f(E_{\p}^i) + f(E_{\p+\q}^j)-1} {\omega + i 0 -(E_{\p}^i + E_{\p+\q}^j)} \, ,
\end{align}
where the coherence factors determined by Eq.~\eqref{Coherence}
\begin{align}\label{app:Coherence}
 C^{(pp)}_{ij;\p,\q}  =1-\frac{\vare_{\p}^i} {E_{\p}^i} + \frac{\vare_{\p+\q}^j} {E_{\p+\q}^j} -
 \frac{\vare_{\p}^i \vare_{\p+\q}^j + \Delta_{\p}^i
 \Delta_{\p+\q}^j} {E_{\p}^i E_{\p+\q}^j} \, .
 \end{align}
Here we assume $\Delta^i_{\p} = - \Delta^j_{\p + \q} = \Delta >0$, and limit the discussion to $\q$ such that the two normal state Fermi surfaces, $\varepsilon^i_{\p} =0$ and $\varepsilon^j_{\p+\q} =0$ have common points in BZ, i.e. they cross and/or touch.
We focus on the singularity at $\omega = 2 \Delta$ which is the lower threshold of quasi-particle excitations and obtain the most singular part of $\mathrm{Im} \chi^0$ at $\omega \rightarrow 2 \Delta$.
In this limit, the momenta contributing to $\mathrm{Im}\chi^0$ are close to the intersection of the original and shifted Fermi surfaces.
For the most singular part of $\mathrm{Im}\chi^0$ we therefore have, $C^{(pp)}_{ij;\p,\q} \approx 2$, and
\begin{align}\label{appB}
\mathrm{Im} \chi^0_{ij}  (\q,\omega) \approx \frac {\pi}{2}\sum_{\p}
\delta\left( \omega  - E_{\p}^i -  E_{\p+\q}^j \right) \, .
\end{align}
Furthermore, at $\omega \rightarrow 2 \Delta$ we expand,
\begin{align}\label{appBa}
E_{\p}^i \approx \Delta + \frac{[\varepsilon_{\p}^i]^2}{ 2 \Delta} \, ,
\end{align}
and with notation $\bar{\omega} =( \omega - 2 \Delta) 2 \Delta$ we rewrite Eq.~\eqref{appB} as
\begin{align}\label{appBb}
\mathrm{Im} \chi^0_{ij}  (\q,\omega) \approx  \pi \Delta \sum_{\p}
\delta\left( \bar{\omega}   - [\varepsilon_{\p}^i]^2 -  [\varepsilon_{\p+\q}^j]^2 \right) \, .
\end{align}
We start with the case when the two Fermi surfaces, $\varepsilon^i_{\p} =0$ and $\varepsilon^i_{\p+\q} =0$ touch.
For the moment we also neglect the dispersion in $p_z$ direction.
We choose the axis frame so that the Fermi velocities at the two Fermi surfaces at the touching point are
$\Vec{v}_{1,2}= v_{1,2} \hat{x}$.
For internal (external) touching of the two Fermi surfaces $\mathrm{sgn}( v_1) =  \pm \mathrm{sgn}  (v_2)$.
Close to the touching point(s),
\begin{align}\label{expand}
\varepsilon_{\p}^i  \approx v_{1,2} p_x + \frac{ p_{y}^2}{ 2 m_{1,2}}\, .
\end{align}
where the momentum is counted from the touching(s) points.
We note that  $m_i >(<) 0$ in Eq.~\eqref{expand} for electron or hole like pockets respectively.
It is convenient to change to new variables,
\begin{align}\label{change}
\xi = \varepsilon_{\p}^1\, , \,\,\, \eta = \varepsilon_{\p}^2\, .
\end{align}
Relation \eqref{change} with the dispersion relation Eq.~\eqref{expand} can be inverted, provided $m_1 v_1 \neq m_2 v_2$ which is a generic situation and $\mathrm{sgn}( v_2 \xi - v_1 \eta) = \mathrm{sgn}( m_2 v_2 - m_1 v_1 )$ as follows
\begin{align}\label{invert}
p_x = \frac{ m_1 \xi - m_2 \eta }{ m_1 v_1 - m_2 v_2 }\, , \,\,\,
p_y = \sqrt{ \frac{v_2 \xi - v_1 \eta }{v_2/ 2 m_1 - v_1 / 2 m_2}   }\, .
\end{align}
We set without loss of generality $v_2 m_2 - v_1 m_1 > 0$, then
\begin{align}\label{appB1}
\mathrm{Im} \chi^0_{ij}  (\q,\omega) \approx  \pi \Delta
\int_{D} \frac{d \xi d \eta}{ (2 \pi)^2} J( \xi,\eta ) \delta\left( \bar{\omega} - \xi^2 - \eta^2 \right)\, ,
\end{align}
where the integration domain, $D$ is $v_2 \xi > v_1 \eta$ and the Jacobian is easily evaluated
\begin{align}\label{J}
J = \left[ 2 \left(\frac{ v_1 }{ m_1 } - \frac{ v_2 }{ m_2 } \right)
( v_2 \xi - v_1 \eta ) \right]^{-1/2}\, .
\end{align}
We next transform to the polar coordinates
\begin{align}\label{polar}
\xi = \rho \cos \phi\, , \,\,\, \eta = \rho \sin \phi\, .
\end{align}
Writing $v_1 = \sqrt{ v_1^2 + v_2^2 } \cos \phi_0$, $v_2 = \sqrt{ v_1^2 + v_2^2 } \sin \phi_0$, we
have
\begin{align}\label{combination}
v_2 \xi - v_1 \eta = \rho  \sqrt{ v_1^2 + v_2^2 } \sin ( \phi_0 - \phi)\, .
\end{align}
Substituting Eqs.~\eqref{J}, \eqref{polar} and \eqref{combination} in Eq.~\eqref{appB1} we obtain
\begin{align}\label{appB2}
\mathrm{Im}& \chi^0_{ij}  (\q,\omega) \approx  \frac{\Delta}{ 4 \pi }
\left[ 2 \left(\frac{ v_1 }{ m_1 } - \frac{ v_2 }{ m_2 } \right) \sqrt{v_1^2 + v_2^2} \right]^{1/2}
\notag \\
\times & \int_{\phi_0 - \pi }^{\phi_0} \frac{d \phi}{\sqrt{ \sin( \phi - \phi_0 ) }}
\int_0^{\infty} d \rho \sqrt{\rho} \delta\left( \bar{\omega} - \rho^2 \right)\, .
\end{align}
The angular integration is convergent,
\begin{align}\label{phi_int}
\int_{\phi_0 - \pi }^{\phi_0} \frac{d \phi}{\sqrt{ \sin( \phi - \phi_0 ) }} = 2 \sqrt{2} K(1/2)\approx 5.2 \, ,
\end{align}
where $K(x)$ is the complete elliptic integral of the first kind, and the $\rho$ integration trivially gives
\begin{align}\label{rho_int}
\int_0^{\infty} d \rho \sqrt{\rho} \delta\left( \bar{\omega} - \rho^2 \right)
=
\frac{ 1 }{ 2 \bar{\omega}^{1/4} }\, .
\end{align}
In result the singular part at $\omega - 2 \Delta \ll \Delta$ is \cite{Maiti2011a}
\begin{align}\label{appB3}
\mathrm{Im}& \chi^0_{ij}  (\q,\omega) \approx
C \left[\frac{\omega - 2 \Delta}{ 2 \Delta} \right]^{-1/4} \theta( \omega - 2 \Delta)\, ,
\end{align}
where the constant
\begin{align}
C = \frac{K(1/2)}{ 2 \pi }
\left[ 2 \left(\frac{ v_1 }{ m_1 } - \frac{ v_2 }{ m_2 } \right) \frac{\sqrt{v_1^2 + v_2^2}}{ \Delta} \right]^{-1/2} \, .
\end{align}

We now turn to the singularity in $\mathrm{Im} \chi^0$ for three dimensional dispersion relation when the two Fermi surfaces touch.
The possibility of a saddle point touching is not considered here.
We note that the stronger singularity may be obtained in this case.

Instead of \eqref{expand} we have
\begin{align}\label{expand_3D}
\varepsilon_{\p}^{1,2}  \approx v_{1,2} p_x + \frac{ p_{y}^2 + p_z^2 }{ 2 m_{1,2}}\, .
\end{align}
The dispersion anisotropy in the touching, $yz$ plane is expected to play no role and is neglected.
By changing to the polar coordinates in this plane,
\begin{align}
p_y = p_{\perp} \cos \phi\, , \,\,\, p_z = p_{\perp} \sin \phi
\end{align}
we write
\begin{align}\label{perp}
\varepsilon_{\p}^i  \approx v_{1,2} p_x + \frac{ p_{\perp}^2 }{ 2 m_{1,2}}\, .
\end{align}
and \eqref{appB} can be written after a trivial angular integration
\begin{align}\label{appB_3D}
\mathrm{Im} \chi^0_{ij}  (\q,\omega) \approx & \frac{\Delta }{ 4 \pi }
\int_{-\infty}^{\infty} d p_x \int_0^{\infty} d p_{\perp} p_{\perp}
\notag \\
& \times \delta\left( \bar{\omega}   - [\varepsilon_{\p}^i]^2 -  [\varepsilon_{\p+\q}^j]^2 \right) \, .
\end{align}
with $\varepsilon_{\p}^i$ specified by Eq.~\eqref{perp}.
Writing $\int_{0}^{\infty} d p_{\perp} p_{\perp} = (1/2)\int_{-\infty}^{\infty} d p_{\perp} |p_{\perp}|$ we obtain the integral very similar to the two-dimensional case.
Repeating the same steps we arrive at the following expression,
\begin{align}\label{appB_3D_1}
\mathrm{Im} \chi^0_{ij}  (\q,\omega) \approx & \frac{\Delta }{ 8 \pi }
\left| \frac{v_1}{m_2} - \frac{ v_2}{ m_1} \right|^{-1}
\int_{D} d \xi d \eta \delta\left( \bar{\omega} - \xi^2 - \eta^2 \right)\, .
\end{align}
The integral in Eq.~\eqref{appB_3D_1} gives constant,
\begin{align}\label{appB_3D_2}
\int_{D} d \xi d \eta \delta\left( \bar{\omega} - \xi^2 - \eta^2 \right)= \frac{\pi}{2}\, .
\end{align}
Therefore $\mathrm{Im} \chi^0_{ij}  (\q,\omega)$ has a jump discontinuity at $\omega = 2 \Delta$,
\begin{align}\label{appB_3D_3}
\mathrm{Im} \chi^0_{ij}  (\q,\omega) = C' \theta( \omega - 2 \Delta)
\end{align}
with a constant
\begin{align}
C' = \frac{\Delta }{ 16 }
\left| \frac{v_1}{m_2} - \frac{ v_2}{ m_1} \right|^{-1}\, .
\end{align}
Correspondingly the real part of the susceptibility has a logarithmic singularity at $\omega = 2 \Delta$,
\begin{align}\label{KK}
\mathrm{Re} \chi^0_{ij}  (\q,\omega) = \frac{C'}{\pi} \log\left|\frac{E_F}{2 \Delta - \omega} \right|
\end{align}
as follows from the Kramers-Kronig
relations.

While in two dimensions the singularity is algebraic, Eq.~\eqref{appB3}, in three dimensions it is only logarithmic, Eq.~\eqref{KK}.
For that reason the binding energy while at maximum close to touching condition is still exponentially small.
For the ``squarish'' dispersion considered in Ref.~[\onlinecite{Maier2012}] the conditions for the resonance are more favorable because the quasi-one-dimensional dispersion yields strong inverse square root singularity at a $2 \Delta$ threshold.
Moreover the external touching gives stronger resonance.
This observation is limited to the quasi-one-dimensional dispersion.
The singular part of a bare susceptibility is approximately the same for both external and internal touching conditions.
However, the non-singular part originating from the states not influenced by the superconductivity has a large logarithm, $\sim \log(E_F/ \Delta)$ which is a famous $2 k_F$ singularity of a Lindhard function cut by $\Delta$, (here $E_F$ and $k_F$ are Fermi energy and momentum respectively).
Since in higher dimensions Lindhard function is singular but finite at $2k_F$, we, in general, do not expect the external touching to yield a stronger resonance than the internal one.
Nevertheless even in a three dimensional case considered here the binding energy is at local maximum when the touching is external.

\end{appendix}

\end{document}